\documentclass[12pt, draftclsnofoot, onecolumn]{IEEEtran}

\usepackage{subfigure}
\usepackage{url}
\usepackage[utf8]{inputenc}

\ifCLASSINFOpdf
\usepackage[pdftex]{graphicx}
\else
\usepackage[dvips]{graphicx}
\fi

\usepackage[utf8]{inputenc}
\usepackage{epsfig}
\usepackage{epstopdf}
\usepackage{graphicx}
\usepackage{times}
\usepackage{url}
\usepackage{multirow}
\usepackage{multicol}
\usepackage{algorithm}
\usepackage{amsmath,amsthm}
\usepackage{amsfonts,amssymb}
\usepackage{subfigure}
\usepackage{tabularx}
\usepackage{comment}
\usepackage{xspace}
\usepackage{soul}
\usepackage{color}
\usepackage{booktabs}
\usepackage{algpseudocode}
\usepackage{enumerate}
\usepackage{lipsum}
\usepackage[colorinlistoftodos]{todonotes}
\usepackage{float}

\begin{document}
	\title{Capture Aware Sequential Waterfilling for LoraWAN Adaptive Data Rate}
	
	\author{Giuseppe Bianchi\IEEEauthorrefmark{2}, \IEEEauthorblockN{Francesca Cuomo\IEEEauthorrefmark{1}, Domenico Garlisi\IEEEauthorrefmark{3}, Ilenia Tinnirello\IEEEauthorrefmark{3}\\
		\IEEEauthorblockN{\IEEEauthorrefmark{1}CNIT/ University of Rome "La Sapienza", Italy \IEEEauthorrefmark{2}CNIT/ University of Rome "Tor Vergata", Italy, \IEEEauthorrefmark{2}\\CNIT/ University of Palermo, Italy}}}

	\maketitle

    \thispagestyle{empty} 
    \pagestyle{empty}    
	
	\begin{abstract}
	LoRaWAN (Long Range Wide Area Network) is emerging as an attractive network infrastructure for ultra low power Internet of Things devices. Even if the technology itself is quite mature and specified, the currently deployed wireless resource allocation strategies are still coarse and based on rough heuristics. This paper proposes an innovative ``sequential waterfilling'' strategy for assigning Spreading Factors (SF) to End-Devices (ED). Our design relies on three complementary approaches: i) equalize the \emph{Time-on-Air} of the packets transmitted by the system's EDs in each spreading factor's group; ii) balance the spreading factors across multiple access gateways, and iii) keep into account the channel capture, which our experimental results show to be very substantial in LoRa. While retaining an extremely simple and scalable implementation, this strategy yields a significant improvement (up to 38\%) in the network capacity over the legacy Adaptive Data Rate (ADR), and appears to be extremely robust to different operating/load conditions and network topology configurations.
	\end{abstract}
	
	
\section{Introduction}
The Internet of Things (IoT) community is currently focusing on the design of large (city/regional-scale) network infrastructures via either proprietary technologies such as LoRaWAN \cite{LoRa} or 3GPP standards like Narrow-Band-IoT (NB-IoT, \cite{NBIoT}). In this paper we specifically focus on LoRaWAN, a promising solution for large-scale ultra low power IoT deployments  \cite{LoraforIoT}\cite{Alliance}. By operating in the unlicensed Industrial, Scientific and Medical (ISM) radio bands, LoRaWAN's ease of deployment makes such a technology a serious candidate for revolutionizing pervasive smart-city services in fields such as transportation, energy or health \cite{Health16}.  

In the rest of the paper we refer to LoRa as for the physical layer protocol and to LoRaWAN for the networking part of the system.

While LoRaWAN inherits several ``classical'' wireless network features, such as native support of multiple transmission rates, it exhibits many peculiar characteristics which make the resource allocation problem quite original and still prone to significant improvements. Indeed, multi-rate support is specifically accomplished in LoRaWAN by exploiting six different {\em spreading factors} for transmitting packets between EDs and network GateWays (GW). The selection of the spreading factor is a compromise between message duration and packet delivery probability (or, dually, communication range). In principle, each node can communicate by selecting the {\em minimum} SF which permits correct reception by an intended gateway; indeed, this is the design target of the legacy ADR strategy currently employed in LoRaWAN deployments \cite{Alliance13}. However, as duly discussed in this paper, three further aspects can be considered to improve the total network capacity. A detailed analysis of the impact of different configurable parameters on the performance of the legacy ADR mechanism that runs on both the EDs and the network is presented in \cite{Agile}.

First, the main contribution of interference is among nodes transmitting with the same spreading factor. A second and quite specific feature of LoRa is the extent to which the so-called channel ``capture'' {\em may} enter into play. Obviously, capture occurs in any wireless technology. Whenever two signals are simultaneously on the air, provided that the difference in the signal strength is sufficiently large, a receiver may still correctly decode the stronger signal. While previous work has duly experimentally assessed \cite{Bor2016} and mathematically modeled \cite{Bankov17} the quantitative impact of capture in LoRa, we are not aware of previous work that {\em constructively} exploits the capture for resource allocation. As a matter of fact, as discussed in details later on, packet capture in LoRa is {\em very} significant: the robust form of frequency modulation employed in LoRa brings about the possibility to capture a packet transmission even for signal strength differences in the order of as little as 1 dB. Goal of this paper is (also) to exploit such distinctive LoRa's feature, so far apparently neglected by prior resource allocation works. Third, and in contrast to many classical wireless local area technologies where Access Point (AP) selection is explicit (e.g. performed by means of an association procedure), LoRaWAN does not mandate a link-level association to a specific radio access station (named ``gateway'' in LoRaWAN's jargon), and thus a same signal may be seamlessly received by multiple GWs, even tens of kilometers away when large spreading factors are employed. As shown in this paper, this fact, once explicitly accounted for, may bring about significant gains with respect to resource allocation strategies, such as ADR \cite{ADR17}, which are ``just'' designed to optimize the transmission towards a target (closer) gateway. 

The main contribution of this paper consists in the design of a LoRaWAN resource allocation scheme which jointly and constructively takes into account all the three specific aforementioned features. Scheme's first baseline principle, rigorously proven via a mathematical model (see section \ref{Sec:optimal}), consists in attempting to assign different spreading factors to EDs in order to equalize the {\em Time-on-Air (ToA)} for each spreading factor's group. The SF allocation is also performed taking in to account the nodes Received Signal Strength Indicator ($RSSI$) and the network topology, in order to optimize the channel capture and improve the performance. We show the advantages of the proposed scheme by simulations, where the simulations parameters are extracted by real network deployments. Our proposed approach, named EXPLoRa-Capture (EXPLoRa-C), is a significant evolution of the algorithm presented in our previous work \cite{Explora}, named EXPLoRa, that was focused only on single gateway and no capture. In this previous work the scheme is intuitively postulated.
In details, the contributions of the paper are.
\begin{enumerate}
\item we analytically provide the optimal load allocation across SFs under both the assumptions of perfect or imperfect orthogonality among different SFs and evaluate the LoRaWAN cell capacity in presence of channel capture effect;
\item we propose and design EXPLoRa-C, a resource allocation strategy that takes advantage the SF balancing, the channel capture effect and the network topology to improve the network performance; 
\item we perform an extensive performance analysis both in a single gateway scenario and in a multi-gateway one even with real world data (268 LoRaWAN water meters);
\item we consider the possibility that different network operators exist in the area covered by multiple GWs when one or more not use the proposed allocation scheme (external interference network).
\end{enumerate}

The rest of the paper is organized as follows. The background on how LoRaWAN operates is provided in Section \ref{Sec:background}.
Section \ref{sec:capacity} analyses the capacity that can be achieved in a LoRa cell and derives the optimal load allocation on the basis of the EXPLoRa paradigm (see Sec. \ref{Sec:optimal}). The capacity improvements under the capture effects are analytically discussed in Section \ref{Sec:capture}. The heuristics to implement EXPLoRa-C in both the single cell scenario and multi-gateway one and under the capture effects are in Section \ref{Sec:Explora} while the performance evaluation is presented in Section \ref{sec:performance_evaluation}. Section \ref{sec:related} presents the main works in the current literature while conclusions and future work are discussed in Section \ref{sec:conc}.	

\section{LoRaWAN: background}
\label{Sec:background}
\begin{figure}[t!]
\begin{center}
		\includegraphics[width=100mm]{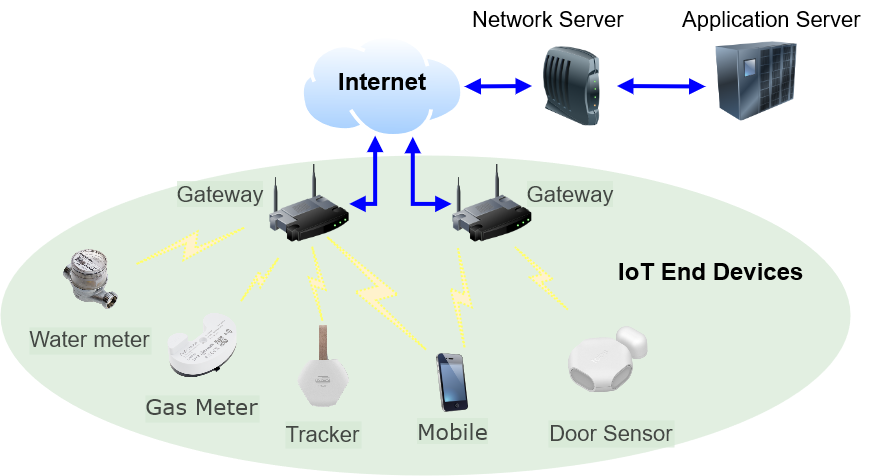}
		\caption{LoRaWAN architecture in case of two gateways with overlapping coverage areas}
		\label{fig:architecture}
\end{center}
\vspace{-6ex}
\end{figure}

The LoRaWAN architecture has a star topology where several EDs are wirelessly interconnected to GWs (see figure \ref{fig:architecture}). A packet transmitted by an ED can be in principle received by multiple far-away GWs, that in turn forward the collected packets to a Network Server (NS) interacting with the Application Server (AS). Battery-powered LoRa EDs are meant to last for a long time (in some scenarios, up to 10 years or even more); as such they communicate using a very low power and low bit rate. Still, they are able to reach quite long distances owing to very robust signal spreading techniques. Communication is bi-directional, although uplink communications from EDs to the NS are strongly favoured.
	
LoRa operates in the unlicensed ISM radio band that are available worldwide. In Europe, it uses the ISM frequencies in the range [$863 MHz - 870 MHz$]. While LoRa is a proprietary technology developed by Semtech \cite{LoRa}, LoRaWAN specification are publicly available and promoted by the open-source LoRa Alliance \cite{Alliance13}. 
LoRa transmissions are regulated by having a maximum transmission power to $25\;mW$ ($14\;dBm$) in the uplink and maximum transmission power of $0.5\;W$ ($27\;dBm$) in the downlink.
Moreover, it employs a duty cycle of $0.1\%$, $1.0\%$ and $10\%$ per day, depending on the channel.
Communication between EDs and GWs is spread out on different frequency channels and data rates. LoRa uses up to 6 different programmable SF: 7, 8, 9, 10, 11, 12. Furthermore, also the adopted bandwidth can be configured: $125\:kHz$, $250\:kHz$ and $500\:kHz$ (typically $125\:kHz$ for the $868$ ISM band). For a given SF, the narrower the bandwidth, the higher the receiver sensitivity.
LoRa is a chirp spread spectrum modulation, which uses frequency chirps with a linear variation of frequency over time in order to encode information. This modulation is immune to the doppler effect and also quite cheap to be implemented. It offers a sensitivity of the order of $-130\;dBm$.
The LoRa Data Rate ($DR$) depends on the Bandwidth ($BW$) in Hz, the spreading factor $sf$ and the Coding Rate ($CR$) as:
    \begin{equation}
    \label{eq:datarate}
        DR=sf\cdot \frac{BW}{2^{sf}}\cdot CR
    \end{equation}
where the symbols/sec are given by $BW/2^{sf}$ with $sf\in\{7-12\}$ and the channel coding rate $CR$ is $4/(4+RDD)$ with the number of redundancy bits $RDD= 1,\cdots,4$.\\
The symbol duration (sec) is calculated as follow:
\begin{equation}
    \label{eq:sym}
        T_{sym}= 2^{sf}/BW.
\end{equation}

LoRa devices use a high spreading factor when the signal is weak or there is a strong interference in the used channel. If an ED is far away from a gateway, the signal gets weaker and therefore needs a higher spreading factor. Using a high SF means a longer symbol duration so a longer $ToA$, i.e., the total transmission time of a LoRa packet. 
The selection of the data rate is a trade-off between communication range and packet duration. Packets transmitted with different SFs, in principle, generate few interference with each other. To maximize both battery life of the EDs and overall network capacity, the LoRaWAN can manage the data rate and RF output for each ED individually by means of an ADR scheme \cite{ADR17}. This mechanism determines the transmission parameters (SF and transmit power) of the ED based on the estimation of the link budget in the uplink and the threshold of the Signal to Interference Ratio ($SIR$) for decoding the packet correctly at the current data rate. 
When the legacy LoRaWAN ADR is enabled, the network will be optimized to use the fastest data rate possible for each ED.

\section{Capacity of LoRa Cells}
\label{sec:capacity}
LoRa cells work as non-slotted Aloha systems. Under Poisson packet arrivals, it is possible to simply evaluate the cell throughput as $G \cdot e^{-2G}$, being $G$ the normalized load offered to the cell. The probability of correctly receiving a packet transmission, which is a typical metric considered for characterizing LoRaWAN systems (often called Data Extraction Rate - DER) is given by $e^{-2G}$. 
Since different orthogonal spreading factors are available, the system works as the super-position of multiple coexisting (but independent) Aloha systems, each one experiencing the load due to the nodes employing a given spreading factor.  

Let $n_{sf}$ be the total number of EDs in the cell employing a SF equal to $sf$ (with $sf\in\{7, 12\}$). The time interval required for transmitting a packet is given by the sum of the preamble time, which lasts $m_{ph}$ symbol times $T_{sym}$ as in \eqref{eq:sym}, and payload transmission time. Since each symbol time codes $sf$ bits and a channel coding with rate $CR = 4/(4+RDD)$ is applied, the time $ToA_{sf}$ required for transmitting over the air a packet long $P$ bytes with spreading factor $sf$ can be expressed, for simplicity, as:
 \begin{equation}
    \label{eq:toa}
    ToA_{sf} =(m_{ph} + \lceil \frac{8 P}{4 sf} \rceil \cdot (4 + RDD))\cdot T_{sym}
\end{equation}


Assuming that every ED generates packets with a source rate of $s$ pkt/s, the normalized load using spreading factor $sf$ can be expressed as $ G_{sf} = n_{sf} \cdot s \cdot ToA_{sf}.$ The total cell capacity results equal to $\sum_{k=7}^{12} G_{k} e^{-2 G_{k}}$ and can dramatically decrease (down to zero) as the loads $G_{sf}$ increase.

Figure \ref{fig:optimal}-a compares the $DER$ performance of an Aloha system (lines) with the results obtained by simulating a LoRa cell (points) as a function of the offered load. Simulations have been obtained by using the public available LoRaSim simulator \cite{LoraSim}, while the offered load is expressed by the number of EDs transmitting in the cell at a source rate of 1 packet every $90sec$, with a packet size of 20 bytes. All the nodes are configured with the same SF and different curves refer to settings which vary from SF 7 to SF 12. From the figure, we can notice that the Aloha model well describes the system behavior. The performance degrade significantly using high spreading factors because larger $ToA$s correspond to higher load conditions. For instance, with 500 EDs, the $DER$ is almost zero for $sf=11$ or $sf=12$, while it is still above 0.5 for $sf=7$.

\subsection{Optimal load allocations across SFs}
\label{Sec:optimal}

\begin{figure}[t!]
\begin{center}
		\subfigure[]{\includegraphics[width=0.42\columnwidth]{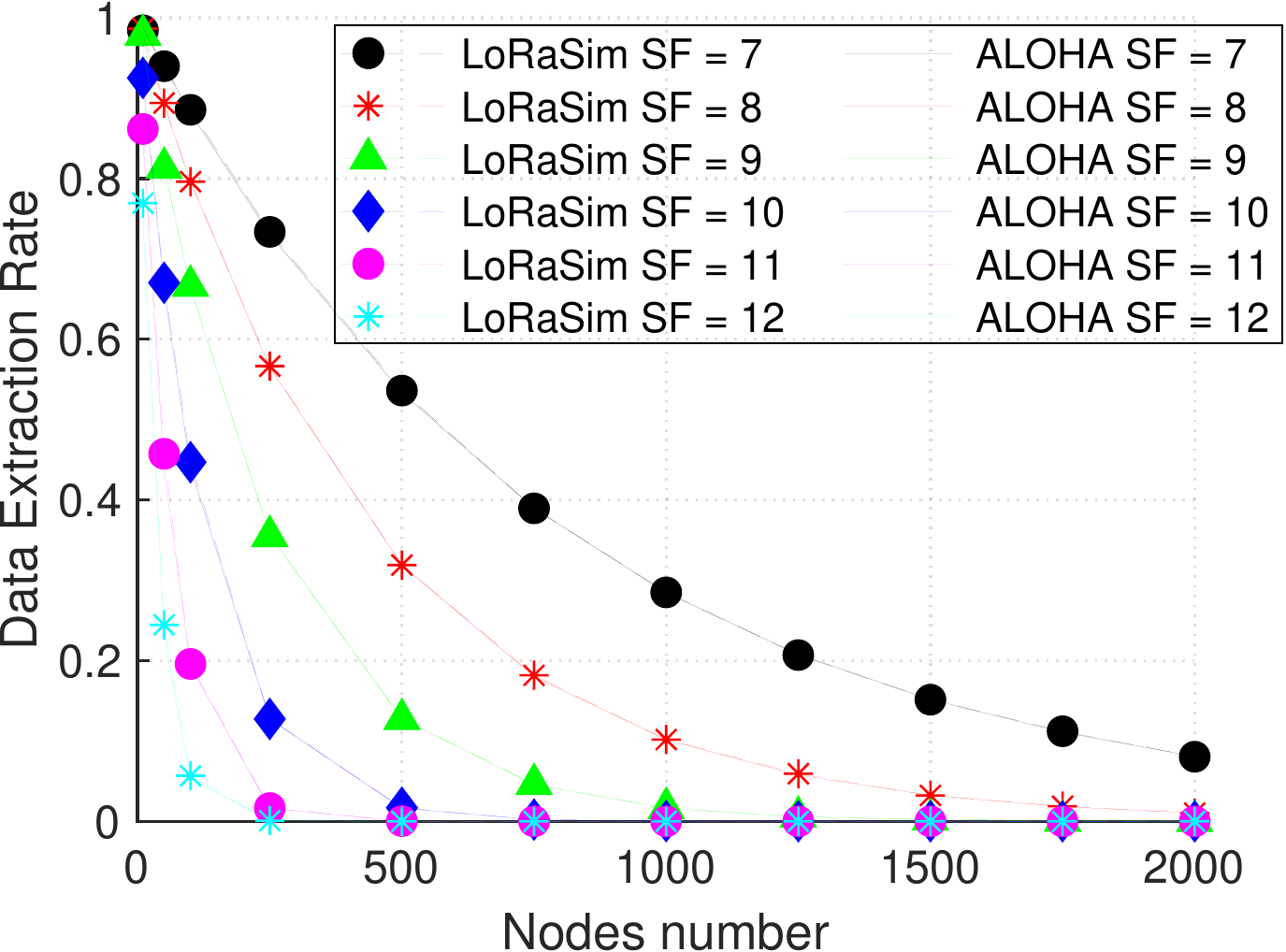}\label{fig:aloha1}}
		\subfigure[]{\includegraphics[width=0.42\columnwidth]{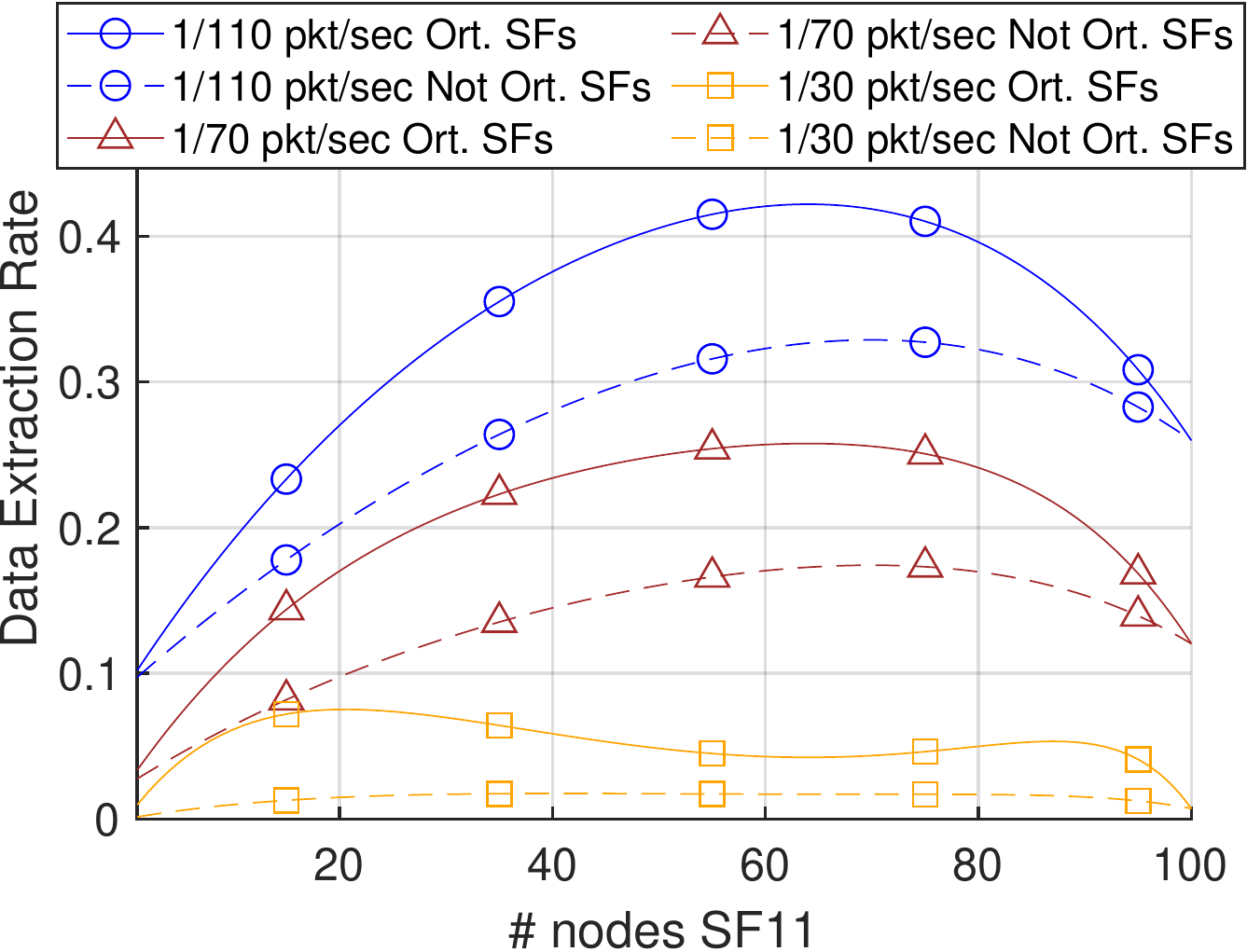}\label{fig:optimal1}}
		\subfigure[]{\includegraphics[width=0.42\columnwidth]{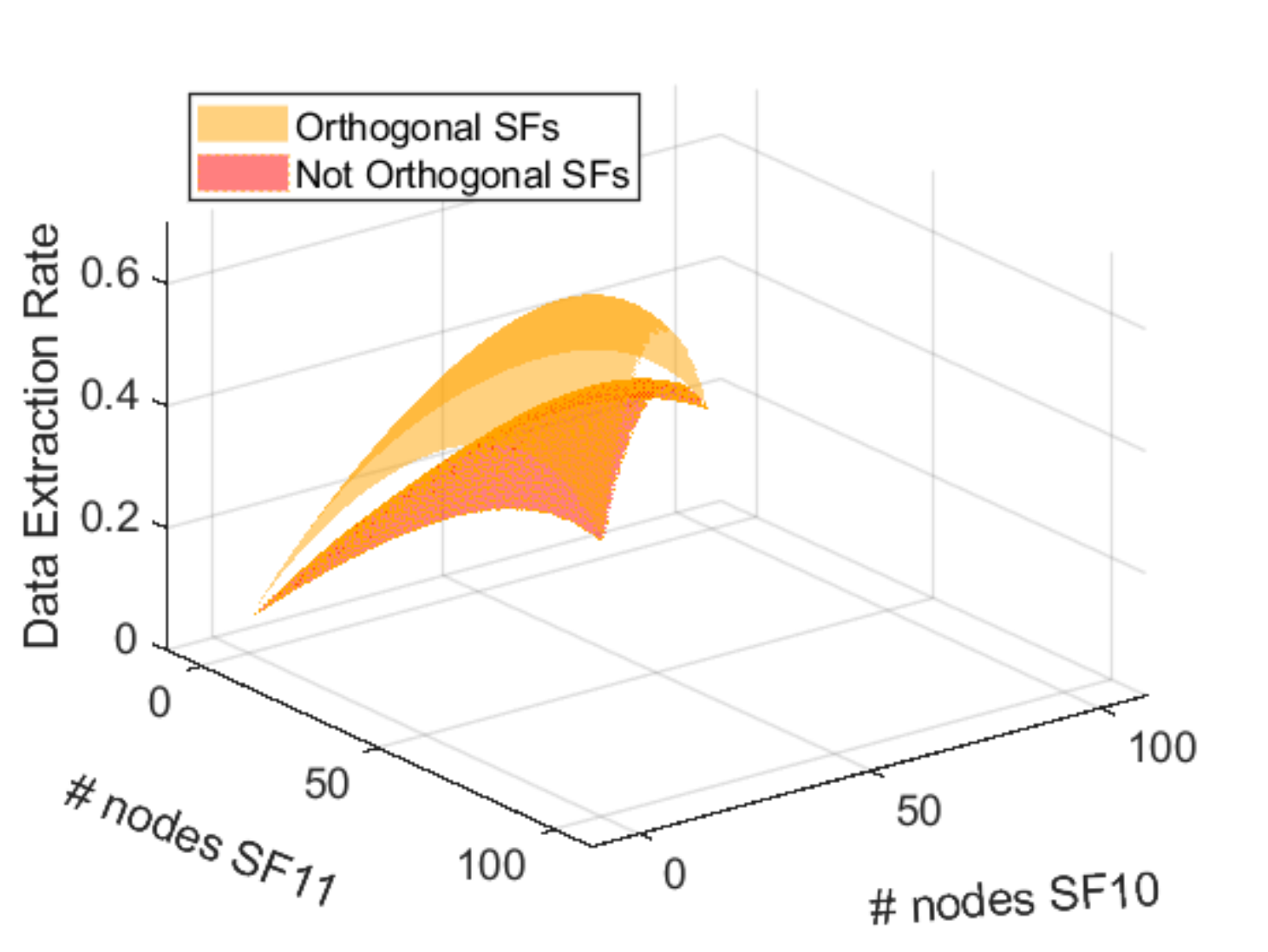}\label{fig:optimal2}}
		\caption{Data Extraction Rate as a function of the number of EDs configured on each SF. When single SF is used in comparison with Aloha formula (a). When only 2 SFs are used, $sf=11$ and $sf=12$ (b) and when 3 SFs are used $sf=10$, $sf=11$ and $sf=12$ (c).}
	\label{fig:optimal}
\end{center}
\vspace{-6ex}
\end{figure}

The problem of spreading factor allocation in a network with $N$ total EDs can be modeled by the choice of $n_7$, $n_8$, $\cdots$, $n_{12}$, i.e. the choice of the number of EDs using each available spreading factor, with the constraint that $\sum_{sf} n_{sf}=N$.\\
A possible optimization criterion is to maximize the average data extraction rate:
\begin{equation}
    \label{eq:meanDER}
    E[DER]= \frac{ \sum_{k=7}^{11} n_{k} e^{-2 n_{k} \cdot s \cdot ToA_{k}}}{N}
\end{equation}
If we replace $n_{12}$ with $N-\sum_{k=7}^{11} n_{k}$ and null all the derivatives $\frac{\partial E[DER]}{\partial n_{sf}}$  with respect to a generic $n_{sf}$ with $sf \neq 12$, we obtain:
\begin{align}
 \label{eq:derivateDER}
    e^{-2 n_{sf} \cdot s \cdot ToA_{sf}} (1\!-\!2 n_{sf} \! \cdot \! s \! \cdot \! ToA_{sf}) \! - \! e^{-2 (N \! - \! \sum_{k=7}^{11} n_{k}) \cdot s \cdot ToA_{12}} \cdot 
  [1-2 (N-\sum_{k=7}^{11} n_{k}) \cdot s \cdot ToA_{12}] = 0
\end{align}
    
By relaxing the constraint of an integer number of end devices per each spreading factor, i.e. by permitting real values for each $n_i$, from the previous equation, it is evident that the solution $n_{sf} ToA_{sf} = n_{12} ToA_{12}$ $\forall sf$, can be a maximum or a minimum because it nulls all the derivatives. By solving for all the SFs and considering the constraint on the total number of EDs, we have:
\begin{equation} 
\label{eq:optimal} 
  n_{sf}^*= \frac{ToA_{12}}{ToA_{sf}} \frac{N}{\sum_{{k}=7}^{12} ToA_{12}/ToA_{{k}}} \quad \forall sf
\end{equation}
If we assume a network working in stable conditions and therefore $1-2 n_{sf} \cdot s \cdot ToA_{sf}$ is greater than zero, the solution $\bold{n^*}=[n_7^*, n_8^*, \cdots n_{12}^*]$  is a maximum.
For very high loads, when $1-2 n_{sf} \cdot s \cdot ToA_{sf} < -1$, the solution $\bold{n^*}$ becomes a minimum. In such a condition, it is interesting to see that $E[DER]$ exhibits other maximization solutions, which are obtained by enforcing a normalized offered load equal to 0.5 (i.e. $n_{sf} = \frac{0.5}{s \cdot ToA_{sf}})$ in all the SFs except one $sf=\widehat{sf}$, in which all the residual $N-\sum_{k=7, k \neq \widehat{sf}}^{11} n_{k}$ are allocated. In other words, for high load conditions, $E[DER]$ is maximized by leaving one SF working in unstable conditions (and in particular, the optimal choice is $\widehat{sf}=12$), and by optimizing the load of all the remaining SFs by setting $G_{sf}=0.5$. 
Obviously, since in practice only an integer number of stations are possible, the final solution has to be obtained by rounding $\bold{n}^*$. 

Figure \ref{fig:optimal} visualizes the previous considerations in a system with two (\ref{fig:optimal}-b, continous lines) or three (\ref{fig:optimal}-c, green surface) available SFs and a total number $N$ of EDs equal to 100. When only two SFs are available ($sf=11$ and $sf=12$), the optimal number of nodes configured on each SF can be determined by studying a single variable function. In the figure \ref{fig:optimal}-b we can immediately recognize that the point which nulls the derivative of $DER(n_{11})$ is given by the solution of the equality $n_{11} ToA_{11} = n_{12} ToA_{12}$, that is $n_{11}$=64 nodes, being $ToA_{12}$ about twice as $ToA_{11}$ as results from \eqref{eq:sym} and \eqref{eq:toa}. As the source rate employed by all the nodes increases, the point changes from a maximum to a minimum point (yellow curve). For high load conditions, the optimal choice is to fix the load for one of the two SF and let the other one become congested (that is $n_{11}$=21 nodes in figure \ref{fig:optimal}-b when source rate is 1/30 pkt/sec). Since the number of stations working in stable conditions are maximized when $G_{11}=0.5$ (rather than $G_{12}=0.5$), the global maximum is reached when $n_{11}= 0.5/(s \cdot ToA_{11})$. The figure \ref{fig:optimal}-c shows a 3D plot when an additional SF is considered ($sf=10$), for a source rate $s=1/90 pkt/sec$. Also in this case, the vector $[n_{10}, n_{11}]$ which nulls the derivative can be easily recognized ($n_{10}$=56 and $n_{11}$=28 for the orthogonal SFs scenario). 

\subsection{Effects of inter-SF interference}
In reality, different SFs are not perfectly orthogonal: it may happen that the reception of a target packet transmitted with a given SF is prevented by an overlapping packet transmitted with a different SF, when $SIR$ is lower than a rejection threshold. In \cite{LoRaImperfect}, it has been experimentally shown that the rejection thresholds are almost independent on the SF of the interfering ED and vary in the range between -10dB (for target packets transmitted at SF 7) and -25dB (for target packets transmitted at SF 12). For simplicity, in the following we refer to a constant inter-SF rejection threshold.

Because of imperfect orthogonality, a target ED working on SF $sf$ at a generic distance $r$ will compete not only with the load $G_{sf}=n_{sf} \cdot s \cdot ToA_{sf}$ offered to the same SF, but also with a fraction of the load $G_{-sf}$ working with a SF different from $sf$, corresponding to the EDs closer to the gateway. For a given rejection threshold $SIR$, only EDs placed in a cell sub-region delimited by a radius $\beta \cdot r$, with $\beta=10^{SIR/10\eta}<1$ can interfere with the target ED while transmitting with a different SF. 
Since such a fraction depends on the distance $r$ and since the number of target EDs grows proportionally to $r$ in case of devices uniformly placed within the cell, the average success rate $DER(sf)$ for a generic target ED working on SF $sf$ can be written as:
\begin{equation} 
\label{eq:eq1} 
e^{-2 G_{sf}} \cdot \int_{0}^{R} e^{ - \frac{\beta^2 r^2}{R^2} \sum_{k \neq sf} n_k s \cdot (ToA_k + ToA_{sf}) } \frac{2r}{R} dr
\end{equation}
where each term $e^{ - \frac{\beta^2 r^2}{R^2} n_k s \cdot (ToA_k + ToA_{sf})}$ is the probability that no transmission at SF $k$ has been started in the interval $ToA_k$ before the starting of the target packet, and no other one is originated during the following packet transmission time $ToA_{sf}$. 
It follows:
\begin{equation} 
\label{eq:eq1} 
DER(sf) = e^{-2 G_{sf}} \cdot \frac{1-e^{-\beta^2 \sum_{k \neq sf} n_k \cdot s \cdot (ToA_k + ToA_{sf})} }{ \beta^2 \sum_{k \neq sf} n_k \cdot s \cdot (ToA_k + ToA_{sf}) }
\end{equation}
In stable conditions, when the load offered to each SF is lower than 0.5, such an expression can be approximated as:
\begin{equation} 
DER(sf) = e^{-2 G_{sf}} e^{-\beta^2/2 \sum_kn_k \cdot s \cdot (ToA_k + ToA_{sf})}
\end{equation}
In other words, the total load $L_{sf}$ competing with the target ED is not only $G_{sf}$, but also a fraction $\frac{\beta^2}{2} \frac{ToA_k + ToA_{sf}}{2\cdot ToA_k}$ of the load $G_k$ offered by each different SF $k$ (with $k\neq sf$). 

We can now generalize the previous derivation on SF allocations, by considering as a new optimization criterion the average data extraction rate achieved in presence of inter-SF interference. In such as case, we have:
\begin{multline}
    \label{eq:meanDER2}
    E[DER]=\frac{\sum_{sf=7}^{12} n_{sf} \cdot e^{-2 \cdot L_{sf}}}{N}
    = \frac{ \sum_{sf=7}^{12} n_{sf} e^{-2 n_{sf} \cdot s \cdot ToA_{sf} -\sum_{k \neq sf}^{12} \frac{\beta^2}{2} n_k \cdot s \cdot (ToA_{k}+ToA_{sf})}}{N}
\end{multline}

If we replace again $n_{12}$ with $N-\sum_{l=7}^{11} n_{l}$ and compute the derivatives $\frac{\partial E[DER]}{\partial n_{sf}}$  with respect to a generic $n_{sf}$ with $sf \neq 12$, we obtain:
\begin{multline}
     \label{eq:derivateDER2}
     e^{-2 L_{sf} } [1 - n_{sf} \cdot s (2 \cdot ToA_{sf} - \frac{\beta^2}{2} n_{sf} \cdot (ToA_{12}+ToA_{sf}) ]+ \\ 
    e^{-2 L_k} \sum_{k \neq sf}^{11} \frac{\beta^2}{2} n_k \cdot s \left[ -(ToA_{k}+To  A_{sf})+ 
    (ToA_{k}+ToA_{12}) \right]  +\\
    e^{-2 L_{12} } [-1+ (N-\sum_{l=7}^{11} n_{l} ) \cdot s  ( -\frac{\beta^2}{2} \cdot (ToA_{12}+ToA_{sf}) + 2 \cdot ToA_{12})]
\end{multline}
By permitting real values for each $n_k$, it is easy to show that the vector of unknown allocations $[n_7^*, n_8^* \cdots n_{11}^*]$ for which $L_{k}=L_{sf}=L_{12}$ $\forall k$ nulls all the derivatives $\frac{\partial E[DER]}{\partial n_{sf}}$ with respect to a generic $n_{sf}$. 
Indeed, in such a case we can simplify the exponential terms from the previous expression and note that the sum of all the other terms is equal to $L_{12}-L_{sf}$, which in turns is equal to zero. 
To find the optimal allocations it is required to solve a linear system in the unknown variables $n_7, n_8, \cdots n_{11}$. 
It can be easily shown that each component of the system solution can be written as:
\begin{equation}
n_{sf}^*=\frac{ToA_{12}}{ToA_{sf}} \frac{ N - \frac{\beta^2}{4} N  \left[\sum_{k=7}^{12} (\frac{ToA_{sf}}{ToA_k}-1) + 2\right]}{ \sum_{k=7}^{12} \frac{ToA_{12}}{ToA_k} (1-\frac{\beta^2}{2} )}  \quad \forall sf
\label{e:load2}
\end{equation}
Such an expression coincides with the one derived in the previous section when $\beta$ is equal to zero (i.e. in absence of inter-SF interference).

Figure \ref{fig:optimal} also shows the effects of inter-SF interference on the average $DER$ and on the optimal load allocations across SFs. In particular, the dashed lines in figure \ref{fig:optimal}-b refers to a cell with two SFs, in which the threshold for rejecting inter-SF interference is equal to -10dB and the propagation coefficient $\eta$ is set to 2.9. We can see that the optimal load allocations are now achieved by increasing the number of nodes on SF 11 (from 64 to 70). Similarly, in the red surface in figure \ref{fig:optimal}-c, the optimal load allocations across three SFs is shifted towards an increased number of stations on the highest possible rate. Obviously, the $DER$ achieved under optimal allocations is lower than the one achieved without inter-SF interference.  

\subsection{Capacity Improvements due to channel capture }
\label{Sec:capture}
	\begin{table}[t]
		\caption{Simulation parameters.}
		\label{tab:Simul}
		\centering
		\begin{tabular}[scale=1.8]{lc}
			\toprule
			\textbf{Parameter} & \textbf{Value} \\
			\midrule
			Carrier Frequency (MHz)& $863.0$  \\
			Bandwidth (kHz)& $125$  \\
			Code Rate (CR) & 4/5 \\
			Message size [byte]& 20\\
			Message Period & 1 packet every 90 sec\\
			Number of gateway & [1-25]\\
			Number of nodes&[100-8000]\\
			TXPower&$14\;dBm$\\
			Path loss values& $\eta=2.9$, $\sigma^{2}=0$, $\overline{L_{pl}}(40m)=-66\;dB$\\
			\bottomrule
		\end{tabular}
		\vspace{-6ex}
	\end{table}
	As discussed in \cite{Bor2016}, and experimentally validated in \cite{LoRaImperfect}, LoRa modulation is very robust to Gaussian noise, but also to self-interference due to colliding transmitters. Indeed, in case of collisions between two or more transmitters, a $SIR$ value of just very few dBs (actually, as little as $1\;dB$ for all the SFs in the experiments described in \cite{LoRaImperfect}, versus the $6\;dB$ specified in \cite{LoRa}) is enough for correctly demodulating the strongest colliding signal. This phenomenon, called ``\emph{channel capture}'', has a strong impact on the scalability of LoRa technology, because the deployment of multiple gateways can significantly boost the capture probability and thus the overall network capacity.

A simple approximation of the throughput improvement due to  channel capture  can be obtained by considering that in most practical cases, a target ED collides with a single interfering signal at time. This assumption is reasonable when the cell works in stable conditions and collisions involving multiple overlapping packets are rare or have a dominant contribution in the interfering power. 
Under this approximation, a target ED employing a given spreading factor $sf$ is actually competing with a fraction of the total load $G_{sf}$. 
Indeed, neglecting the effect of random fading and assuming an attenuation law of type $r^{-\eta}$, all the interfering nodes at distance higher than $\alpha \cdot r$, with $\alpha= 10^{SIR/10\eta}>1$, do not prevent the correct demodulation of the target ED.
The smaller the $\alpha$ coefficient, the smaller the real competing load is.
Therefore, the cell throughput in presence of channel capture  can be obtained by generalizing the Aloha results as:
    \begin{align}
       \label{s_capture}
    S_c(G_{sf}) = 2 \pi \int_{0}^{R/\alpha} \delta_{sf} e^{-2 \frac{ \alpha^2 r^2}{R^2} G_{sf}}  r \cdot dr  + \delta_{sf} (\pi R^2 - \pi R^2/\alpha^2) e^{-2 \cdot G_{sf}} 
	\end{align}

where $\delta_{sf}=G_{sf} /(\pi R^2)$ is the density of  load offered to spreading factor $i$ and $R$ the cell radius. 
The $DER$ is simply given  by $S_c(G_{sf})/G_{sf}$.

We quantified the performance results of LoRa cells in presence of channel capture by using both our simplified model and the already mentioned LoRaSim simulator \cite{LoraSim}. 
Unless otherwise specified, table \ref{tab:Simul} shows the scenario parameters used in our simulations, where $\overline{L_{pl}}(d_{0})$ is the mean path loss at a reference distance $d_{0}=40m$.
%
%
%

\begin{figure}[t]
	\begin{center}
	\subfigure[]{\includegraphics[width=0.42\columnwidth]{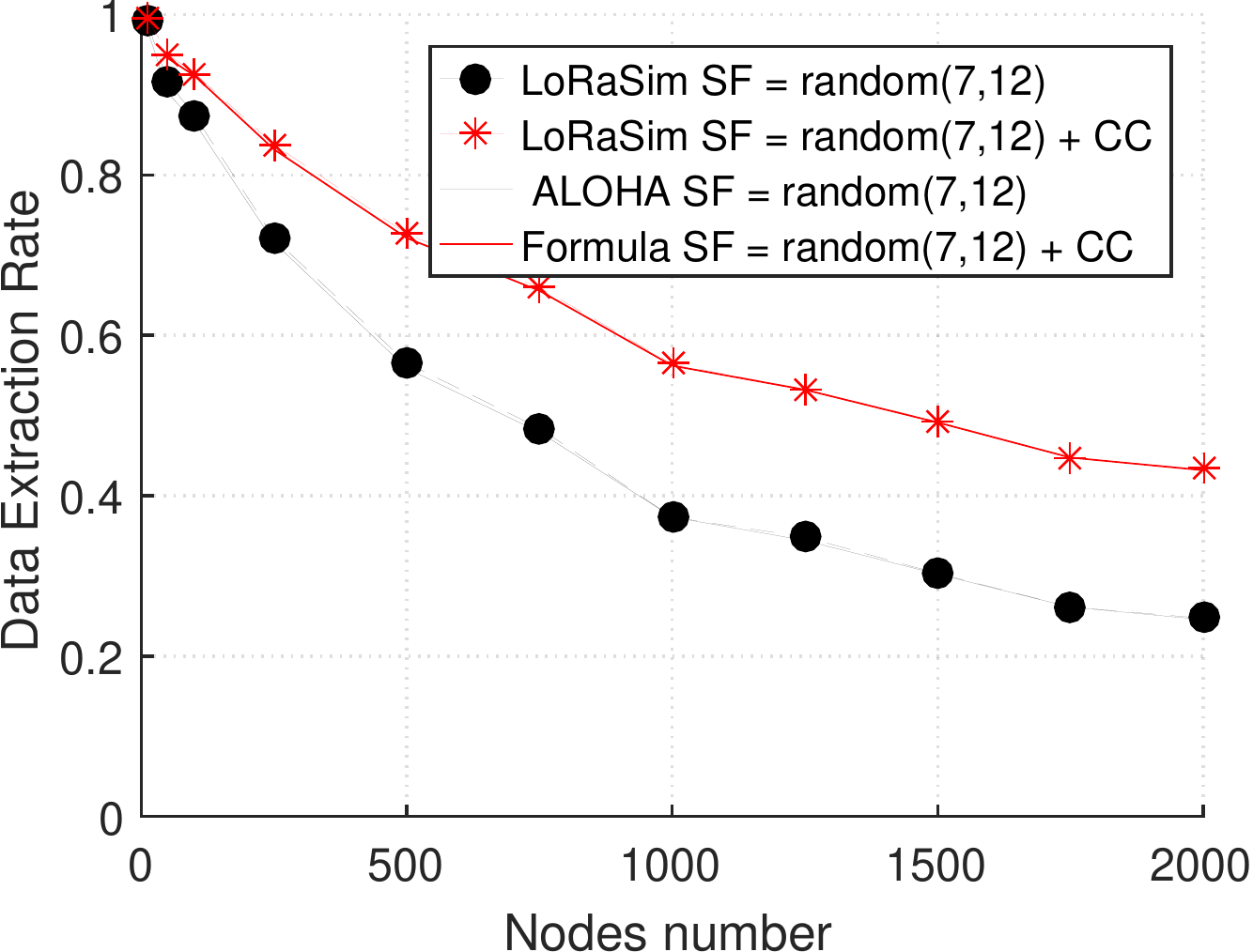}\label{fig:aloha2}}
    \subfigure[]{\includegraphics[width=0.42\columnwidth]{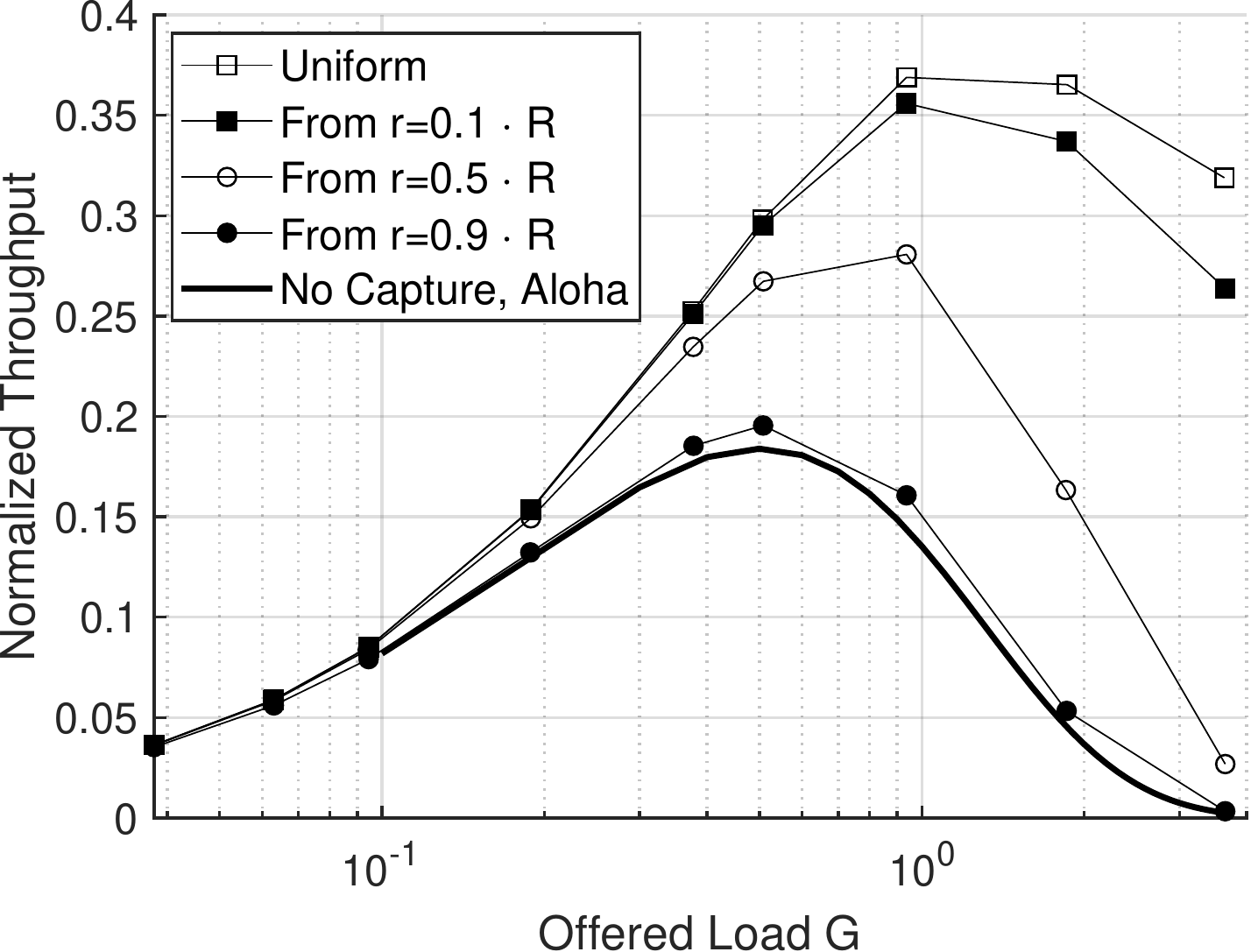}\label{fig:cap1}}
	\subfigure[]{\includegraphics[width=0.42\columnwidth]{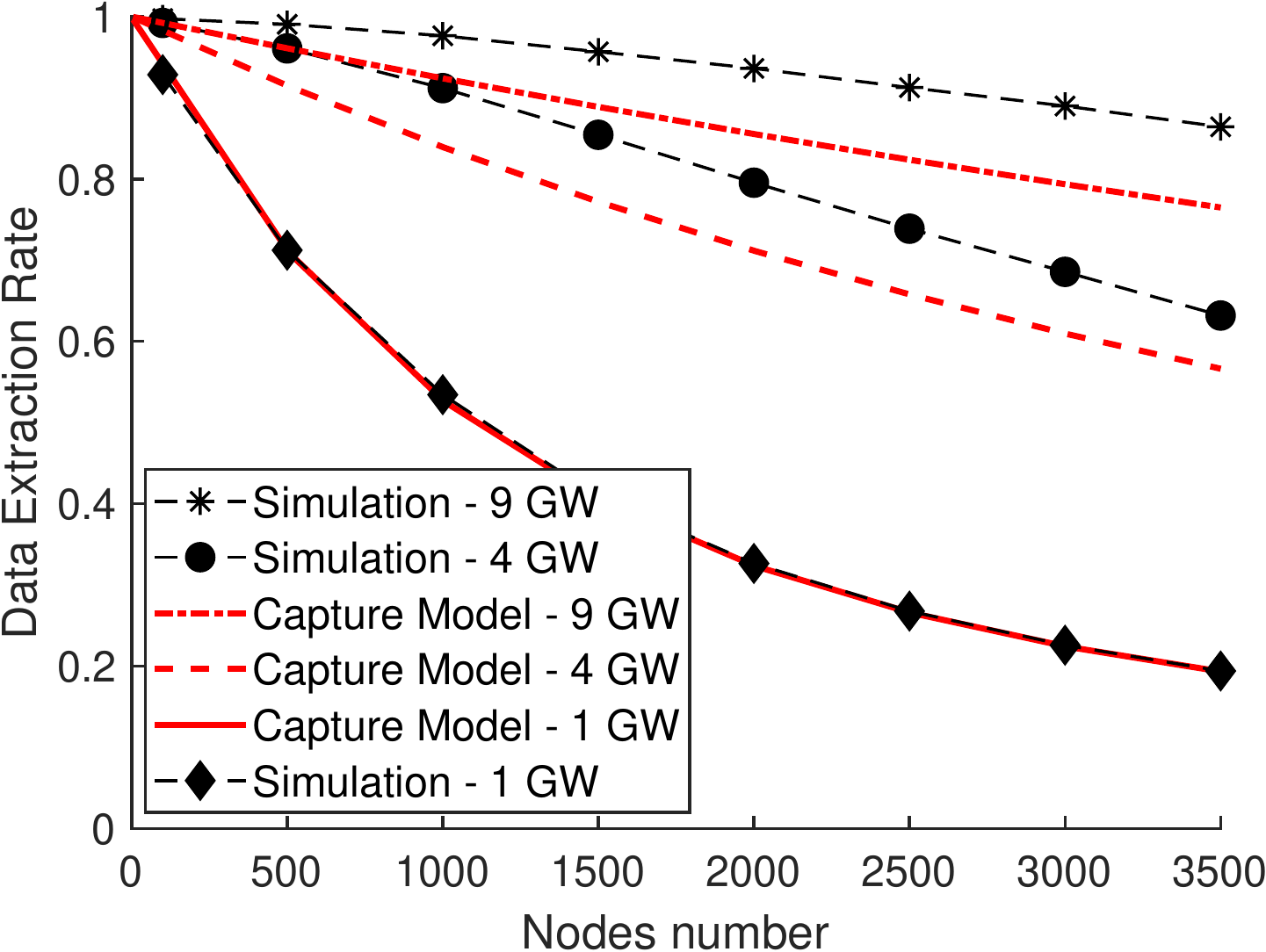}\label{fig:cap2}}
	\caption{
	Data Extraction Rate as a function of the number of EDs uniformly distributed among all the SFs (a), without channel capture  and with channel capture  (+ CC). Capture effect for different node distributions (b) and in presence of multiple gateways (c).}
	\label{fig:cap}
\end{center}
\vspace{-6ex}
\end{figure}

Figure \ref{fig:cap}-a shows the $DER$ achieved with (red curve) and without (black curve) capture effects as a function of the number of EDs in the cell, for a capture $SIR$ threshold of $1\;dB$. Each ED is configured on a randomly chosen SF between 7 and 12. Simulation results are plotted using points, while the analytic results are given by lines. Simulation results match pretty well the upper bound provided by our model, although we ignore accumulation of interference generated by multiple packets. 

We also considered different node distributions within the cell, in order to understand the impact of node placements on the capture probability. 
Figure \ref{fig:cap}-b shows the capacity results when nodes are uniformly distributed (empty squared points) or placed in a circular ring delimited by the cell radius $R$ and a smaller distance $r_0$ (points according with the legend). 
As $r_0$ approaches the cell radius, the capacity tends to the Aloha case (black bold curve), because nodes have more uniform reception powers and the capture probability becomes negligible. 
Finally, figure \ref{fig:cap}-c shows the capacity results in presence of multiple gateways (namely, $M=4$ and $M=9$). Although the general capacity derivation depends on the gateway placements, we can consider a simple approximation in a simplified scenario of limited load conditions. Since a successful reception is more probable at the closest gateway and the capture effect limits the competing load to the circular area around the closest gateway, being $M$ the number of gateways uniformly spaced in the cell and $S_c(G)$ the throughput perceived under load $G$, the total capacity can be by approximated by $M \sum_{sf} S_c(G_{sf}/M)$. The figure (red curves) confirms that the results are not far from the ones obtained by considering $M$ sub-systems with an offered load of $G/M$: the slight increment of $DER$ quantified in simulation in comparison to the proposed approximation is due to the probability of correctly receiving a packet at a gateway different from the closest one.  
\section{The EXPLoRa-Capture strategy}
\label{Sec:Explora}
As previously stated the SFs allocation has an impact on the distance at which the ED can be located and on the robustness of the radio link in presence of fading. On the other side, the $ToA$ spent by packets sent at different SFs can be significantly different (being the ratio between the minimum and the maximum possible $ToA$ about $2^5$). It follows that SF allocation has also an impact on the system load. The position of nodes employing the same SF plays a further crucial role, especially in sight of the somewhat unexpected capability of LoRa to capture and correctly demodulate a signal even in the presence of a significant interference (see detailed discussion and analysis in the previous section \ref{sec:capacity}). All these aspects are jointly accounted for in the EXPLoRa-Capture strategy for SF allocation, which we detail in this section. 

\subsection{EXPLoRa principles: load balancing} 
\label{Sec:Explora-prin}

Our proposed approach starts from the remark that a rate adaptation strategy merely based on link-level budget/measurements, such as the standard ADR defined by the LoRa Alliance \cite{Alliance13}, cannot take advantage of the (quasi) orthogonal nature of different SFs. For an extreme example, if all network devices are very close to the gateway, they will all select $sf=7$, thus congesting such SF, while all the remaining SFs will remain ``empty''. As proposed in our previous work \cite{Explora} a better allocation strategy consists in ``forcing'' some devices to operate with a higher than necessary SF, thus spending a higher $ToA$, but gaining from a better allocation of load among the available SF-induced ``channels''.

In more details, in our previous work \cite{Explora} we proposed two different variants: a basic solution, called EXPLoRa-Spreading Factor (EXPLoRa-SF), and an enhanced approach named EXPLoRa-Air Time (EXPLoRa-AT). The goal of EXPLoRa-SF was to show that performance can increase by distributing users on different spreading factors. Under EXPLoRA-SF the nodes are equally split between SF sub-channels: although some nodes transmit with a $ToA$ higher than necessary, the reduction on the data rate is compensated by the reduction of the interference caused by simultaneous transmissions from the other nodes. With EXPLoRa-AT we introduced a smarter allocation strategy (which he have here now more rigorously supported in section \ref{Sec:optimal} with theoretical arguments): rather than equally splitting the nodes among different SFs, it equally balances the total $ToA$ spent on each SF. For this reason, assuming that all the EDs employ uniform source rates, the number of nodes in each SF follows the proportion reported in table \ref{tab:airtime}, row orthogonal, as justified in section \ref{sec:capacity}-A.
In presence of inter-SF interference, load balancing can be achieved by updating the proportion between the number of EDs that can be allocated on each SF as derived in equation \ref{e:load2}. Table \ref{tab:airtime} summarizes the load balancing results in the case of orthogonal and non-orthogonal SFs. In this last case, the rejection threshold has been configured to $-16dB$. For non-orthogonal SFs, the portion of EDs transmitting at SF 7 increases, while the allocations performed at SF 11 and SF 12 are reduced to almost zero. 

\begin{figure}[t!]
\begin{center}
	\subfigure[]{\includegraphics[width=0.38\columnwidth]{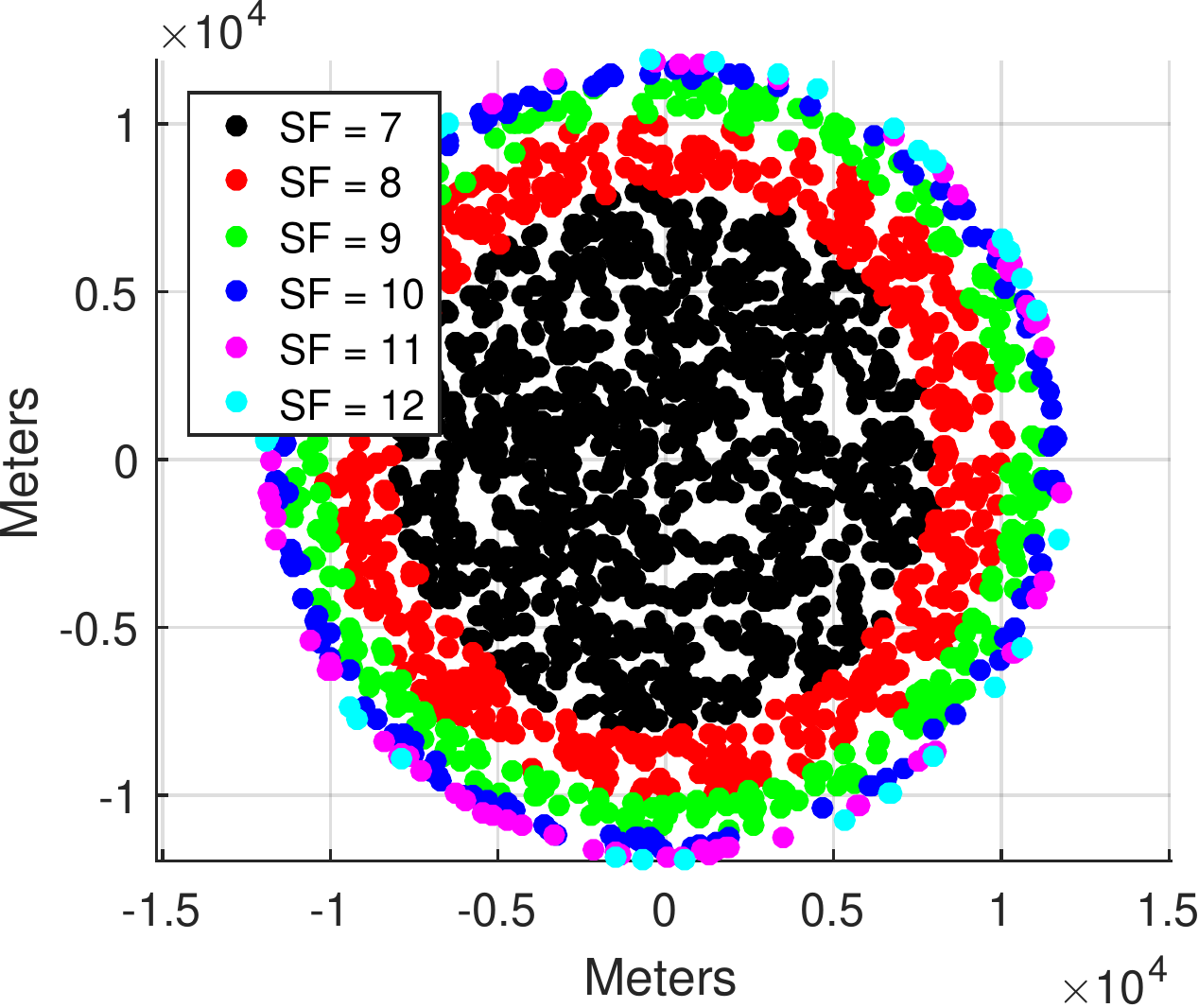}}
	\subfigure[]{\includegraphics[width=0.38\columnwidth]{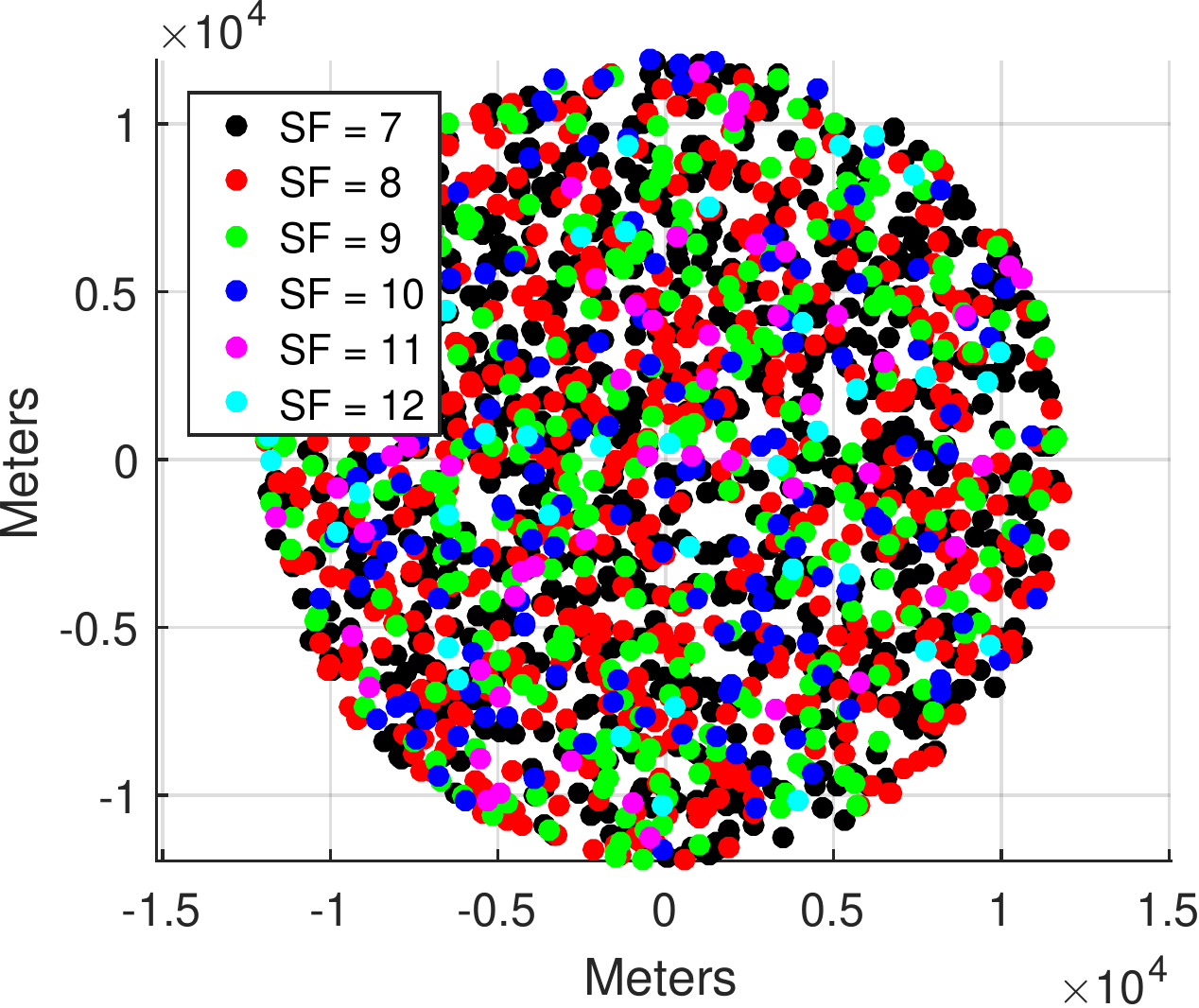}}
	\subfigure[]{\includegraphics[width=0.38\columnwidth]{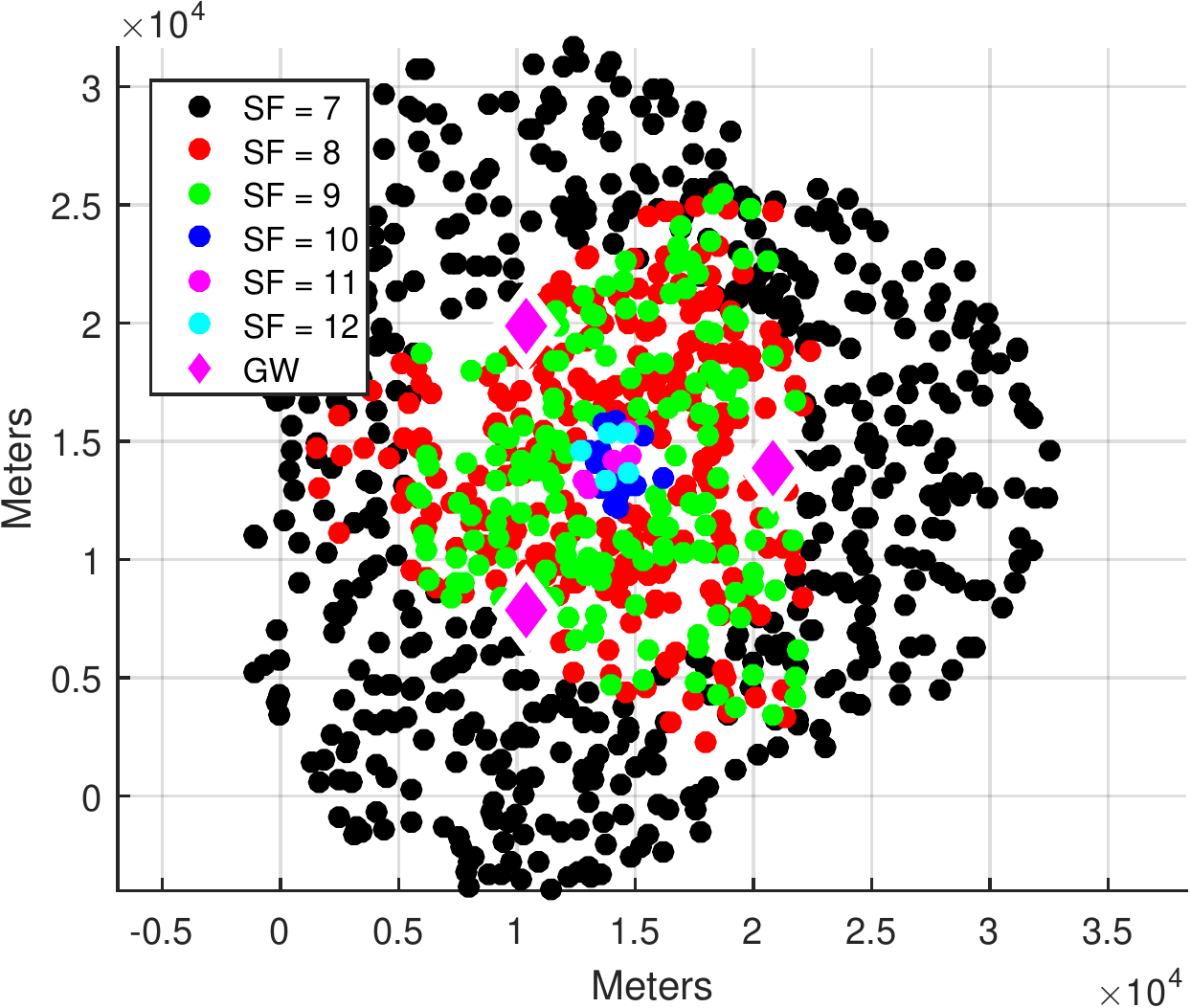}}
\caption{Nodes position and allocated SF with EXPLoRa-AT (a), EXPLoRa-C single gateway (b) and EXPLoRa-C multi gateway (c).}
	\label{fig:sf_explora_node_position_ring}
\end{center}
\vspace{-6ex}
\end{figure}

	\begin{table}[t]
		\centering
		\caption{$ToA$ (in $ms$) as a function of SFs when payload size is 20 byte and coding rate is 4/5; resulting optimal percentages $P_{sf}$ in accordance to optimal load allocation}
		\begin{tabular}{|l|l|l|l|l|l|l|}
			\hline
			\textbf{SF} & 7    & 8   & 9    & 10   & 11   & 12   \\ \hline
			\textbf{$ToA$ [$msec$]} & 49.41&	90.62&	164.86&	329.73&	659.46&	1187.84\\ \hline
			\textbf{$P_{sf}$ [\%], orth.} & 47.02&	25.85&	14.36&	7.18&	3.59&	2.02\\ \textbf{$P_{sf}$ [\%], not orth.} & 50.75&	26.98&	14.07&	0.060 &	0.019 &	0.002\\ \hline
		\end{tabular}
		\label{tab:airtime}
		\vspace{-6ex}
	\end{table}
The effects of load balancing on the proportion of EDs working on the same SF is depicted in figure \ref{fig:sf_explora_node_position_ring}-a for the orthogonal case. For choosing the SF to be used by each ED, we performed sequential allocations in different circular rings, starting from the closest nodes configured with $sf=7$. 
We can easily recognize that about one half of the nodes, colored in black, are using $sf=7$ and all the other EDs follow the optimal $P_{sf}(sf)$ proportions. 
%

%
%
\subsection{EXPLoRa-C details}
\label{Sec:explora-c}
As discussed in section \ref{sec:capacity}, channel capture  can significantly boost the capacity of LoRaWAN, because it is likely that a collision results in the correct reception of the packet received at the highest power, provided that the ratio between the reception power of the two interfering signals is higher than a given $SIR$ threshold. We propose to extend the EXPLoRa scheme in order to maximize the capture probability experienced by the EDs, thus improving the overall system capacity.
The first interesting aspect of the EXPLoRa-C variants is the possibility of implementing the load balance criterion in terms of ``sequential waterfilling''. For facilitating the selection of a data rate compatible with the link budget, EDs are ordered according to their $RSSI$ value (from the highest to the lowest value) and SF allocations are performed sequentially (from the highest rate $sf=7$ to lowest rate $sf=12$). This procedure, different from the previous version, ensures to take the advantage of the channel capture.
The basic idea of this extension, is exploiting the ''spatial'' dimension for reducing the effective load experienced on each SF. For a single cell system, this corresponds to spread as much as possible the EDs working on a given SF within the cell, in order to increase the probability that colliding signals are received with a power ratio higher than the capture threshold. Note that this is generally different from allocating SFs sequentially to the EDs as a function of their ordered $RSSI$ values, because such an allocation could assign the same SF to nodes with similar $RSSI$ values. In other words, EDs employing the same SF should be at different distances from the gateway, rather than concentrated in a circular ring. For a multi-cell system, the spatial dimension can benefit from the availability of multiple gateways: nodes at a similar distance from the closest gateway could indeed be received with very different $RSSI$ values from the neighbor gateways, thus resulting in a good capture probability at a different gateway. 
In order to map these considerations into a new allocation strategy, we introduce the concept of distance between EDs, taking into account both the difference between the $RSSI$ values at the closest gateway, and difference in the set of gateways in their coverage area. The basic idea of EXPLoRa-C is still based on a sequential allocation of SFs, equally sharing the $ToA$ spent at different SFs, but the allocation is performed in multiple rounds, by skipping in each round the decision on nodes which are at a small distance from the previous decision.

{\em Single Cell.} Consider first the case of a single cell with $N$ EDs. The EDs are ordered according to the $RSSI$ value perceived by the cell gateway $GW$, in order to start the decisions on nodes which can employ the highest possible data rate (i.e. $sf=7$). Since high data rates also correspond to shorter transmission times, the number of nodes that can be configured on each SF for equalizing the total $ToA$ is not constant and follows the $P_{sf}$ in table \ref{tab:airtime}. Starting from the first node, EXPLoRa-C assigns the lowest possible SF to each $i$-th ED (we denote with $sf_i$ the SF assigned to the $i$-th ED) till the number of maximum allocations $P_{sf}(sf) \cdot N$ is reached, but only if the distance $RSSI_i - RSSI_j$ (with $j=i-1$ and $i \in [2, N]$) is higher than the capture threshold. Otherwise, node $i$ is left without decision and the next node is processed. After the first allocation round, nodes without decisions are sequentially processed in a second round, in which decisions are randomly taken from the set of SFs which have not reached the total budget of nodes $P_{sf}(sf) \cdot N$. An exemplary allocation following this approach is depicted in figure \ref{fig:sf_explora_node_position_ring}-b in a scenario in which all the SFs can be used even at the cell border. We can see that the same allocation proportions used in figure \ref{fig:sf_explora_node_position_ring}-a are now achieved by spreading the nodes in the whole cell. 

{\em Multi-Gateway.} In case $M$ gateways, $GW_1, GW_2, \dots GW_M$, are deployed in a network with $N$ nodes, EXPLoRa-C is executed $M$ times. Let $N_1, N_2 \dots N_M$ be the number of nodes closest to $GW_1, GW_2, \dots GW_M$, with $N=N_1+N_2+\dots N_M$. 
All the EDs are organized into $M$ sets and ordered as a function of the $RSSI$ value perceived by the closest gateway. For each set, EXPLoRa-C is executed by considering the total number of allocations on each SF as equal to $P_{sf}(sf) \cdot N_m$ (with $m \in [1, M]$). Moreover, the concept of distance between EDs is extended by also considering the neighbor gateways. In case nodes have $RSSI$ values whose difference is lower than the capture threshold but the set of GWs in range is difference, they can still be configured on the same SF. If each $m-th$ gateway $GW_m$ allocates SFs by keeping the $P_{sf}(sf)$ proportion on its nodes $N_m$, the interference generated towards other cells will also respect such a proportion and therefore both the total $ToA$ and the local $ToA$ (i.e. the $ToA$ generated by the closest $N_m$ nodes) seen by each gateway will be balanced.    

{\em Multi-Gateway, multiple network operators.} As a final case, we consider different network operators existing in the area covered by $M$ gateways and some of them do not employ allocations respecting the $P_{sf}$ for their associated devices. In such a case, a mere application of EXPLoRa-C to the EDs under control would not guarantee anymore the load balancing between different SFs. However, it is possible to extend the scheme at each $m$-th gateway, by computing the local $ToA$ $G_{int}(m,sf)$ consumed by the interfering EDs (i.e. different from the $N_m$ set) at each SF, and by deciding the maximum number of allocations on each SF in order to equalize the local $ToA$. 
The total number of interfering EDs seen in cell $m$ is $N_{int}(m)=\sum_{sf} G_{int}(m,sf)/(s \cdot ToA_{sf})$, the NS can evaluate the local $ToA$ at $m$-th gateway because it receives the packets of all EDs in the coverage area (included the interfering EDs). Thus, the optimal value of the possible allocations on each SF in cell $m$ is computed as:
\begin{align}
    n_{sf}^*(m) = max(0, P_{sf}(sf) \cdot (N_m + N_{int}(m)) - G_{int}(m,sf)/(s \cdot ToA_{sf}))
    \label{eq:multi_operator}
\end{align}
where, the sum $N_m + N_{int}(m)$ represents the total number of EDs present in the coverage area of the $m$-th gateway (included the interfering EDs).
In case coverage areas of different operators partially overlap, operators could cooperatively try to avoid allocating the highest possible rate in the overlapping areas, thus reducing the equivalent number of interfering EDs $G_{int}(m,sf)/(s \cdot ToA_{sf})$ for $sf=7$. Such a choice allows the maximization of the number of EDs that each operator can allocate on SF 7, as depicted in the example of figure \ref{fig:sf_explora_node_position_ring}-c.  
%
\begin{algorithm}[t]
    \caption{EXPLoRa-C}
    \label{algo:explora-C}
    \begin{multicols}{2}
		\begin{algorithmic}[1]
				\footnotesize

			\Function{\texttt{EXPLoRa-C}\;}{$ED_{list}$, $\gamma_{th}$, $N$}
    			\State $\forall\;sf\:NUM_{sf}(sf)=0\: sf=\{7-12\}$\Comment{Initialize the number of nodes at the different SFs}
    			\State $ED_{sf}={0}$ \Comment{Initialization of $ED_{sf}$ to 0}
    			\State $ED_{RSSI} \gets$ Sort $ED_{list}$ \Comment{Sort (decreasing order) $ED_{list}$ in accordance to $\overline{RSSI}$}
    			\State $ED_{GW} \gets$  \Comment{Initialization with gateway identification covered by node sorted in accordance to $\overline{RSSI}$}
    			\State Let $P_{sf}(sf)=\{p_7, p_8, p_9, p_{10}, p_{11}, p_{12}\}$ \Comment{Probability distribution function for the EXPLoRA-AT waterfilling as in table \ref{tab:airtime}}

                \State $sf=7$
    			\State $ED_{sf}(1)=sf$\\
    			\vspace{3mm}

    			****PHASE 1*****
    			\For{$i=2$ to $N$}
    			    \If{$ED_{RSSI}(i-1)-ED_{RSSI}(i)>\gamma_{th}$} 
                        \State Assign to $ED_{sf}(i)=sf$
                        \State $NUM_{sf}(sf)=NUM_{sf}(sf)+1$
            			\If{$NUM_{sf}(sf) > {P}_{sf}(sf) \cdot N$}
    			            \State $sf=sf+1$
    			        \EndIf
    			    \EndIf
    			\EndFor\\
                \vspace{3mm}
                ****PHASE 2 (relevant in multi-gateway scanrios)*****
    			\For{$i=2$ to $N$}
    			    \If {$ED_{sf}(i)=0$}
        			    \If{$ED_{GW}(i-1) \neq ED_{GW}(i)$} 
                            \State Assign to $ED_{sf}(i)=sf$
                            \State $NUM_{sf}(sf)=NUM_{sf}(sf)+1$
                			\If{$NUM_{sf}(sf) > {P}_{sf}(sf) \cdot N$}
    			                \State $sf=sf+1$
    			            \EndIf
        			    \EndIf
        			 \EndIf
    			\EndFor\\
    			\vspace{3mm}
    			****PHASE 3*****
    			\State $P_{sf}=P_{sf}-({NUM_{sf}(sf)/N})$ \Comment{Update the distribution function by considering the already assigned nodes to the different SFs}
    			\For{$i=2$ to $N$}
    			    \If {$ED_{sf}(i)=0$}
    			        \State $ED_{sf}(i) \gets \texttt{RAND}_{P_{sf}}\{7-12\}$
    			        \Comment{Select the SF for the node in accordance to the distribution $P_{sf}$}
    			    \EndIf
    			\EndFor
    			\State Return $ED_{sf}$
			\EndFunction
    \end{algorithmic}
    \end{multicols}
\end{algorithm} 
\subsection{EXPLoRa-C algorithm}
The EXPLoRa-C pseudo code for the single-operator case is reported in algorithm \ref{algo:explora-C}.
Let us define a vector $ED_{list}$ of EDs, whose length is $N$ for the single cell scenario or $N_j$ for the multi-cell scenario. The function also requires the $SIR$ threshold $\gamma_{th}$.
The EXPLoRa-C pseudo code works in three different phases, with phase 2 relevant only for the multi-gateway scenario (otherwise only the same gateway is available for all the nodes). 
Starting from $ED_{list}$ and given the $\gamma_{th}$ EXPLoRa-C returns the SF assigned to each node for its transmission. This is provided in a vector, whose length is equal to $ED_{list}$, denoted as $ED_{sf}$.
\begin{figure}[t]
	\centering
	\includegraphics[width=100mm]{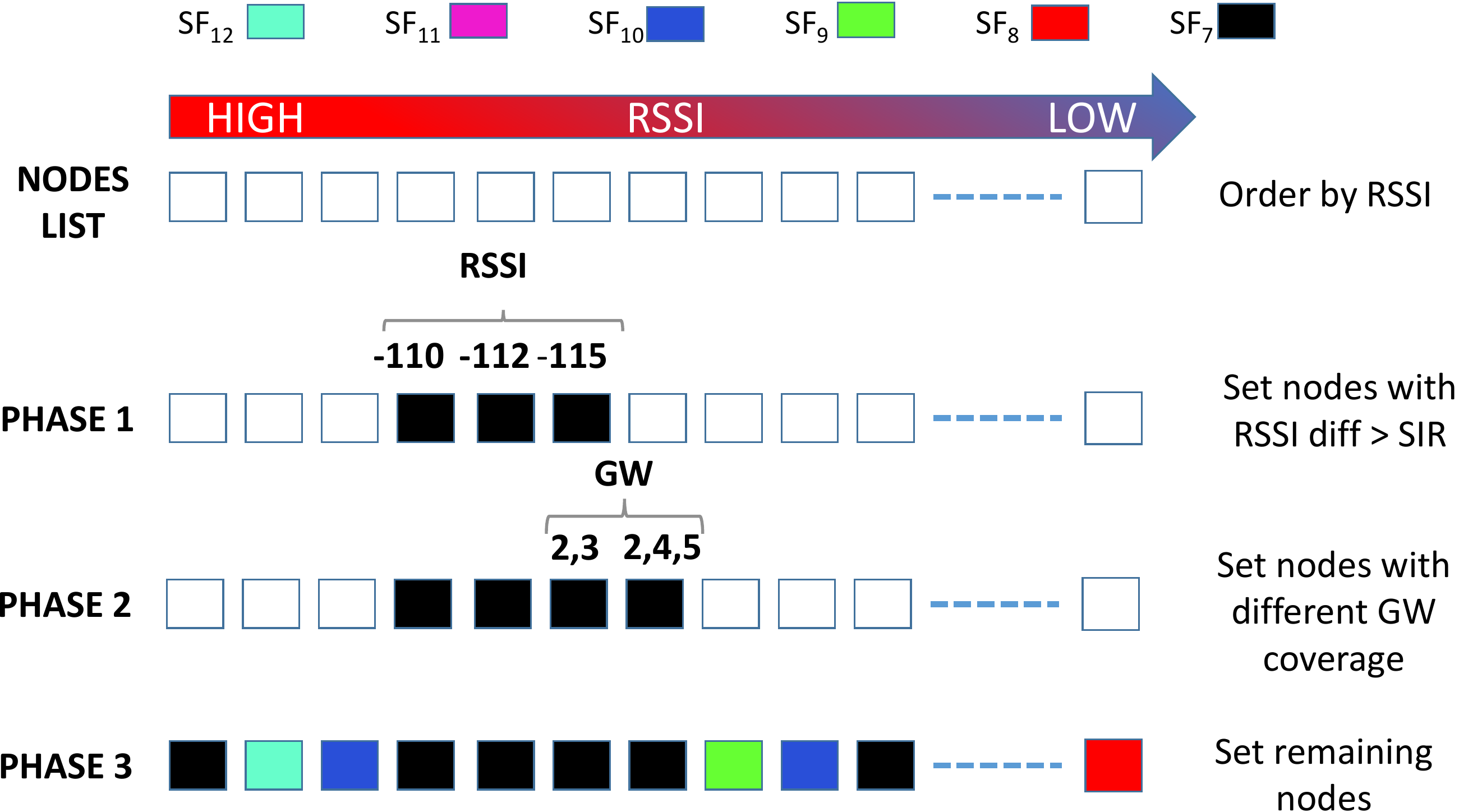}
	\caption{Qualitative representation of the EXPLORA-C mechanism}
	\label{fig:explora-c-figure}
	\vspace{-6ex}
\end{figure}
The first step is the initialization procedure of the algorithm; it sorts nodes in accordance to the measured $\overline{RSSI}$ in decreasing way (from the highest to lower value). This sorting provides a new vector denoted as $ED_{RSSI}$. The parameter $\overline{RSSI}(i)$ is measured as the average of the $RSSI$ values measured by the gateway for the ED $i$ in a time window $W$. 
A qualitative representation of EXPLoRa-C mechanism is in figure \ref{fig:explora-c-figure}, where EDs are represented with squares, empty when the SF has not been assigned, and filled after the SF setting. Different squares colors represent different SFs according with the figure legend. 
In the first phase we derive all the consecutive $ED_{RSSI}$ vector elements whose difference produces a result greater than $\gamma_{th}$, that is assumed in our performance analysis as $1\:dB$. This couple of nodes generate a capture effect due to $RSSI$ values, and they will be set with the same SF value, in the figure, $sf=7$ is assigned to the three consecutive nodes with the respective $RSSI$ values of $-110dB$, $-112dB$ and $-115dB$. 
The algorithm starts by assigning to this couples the $sf=7$ till the number of assigned occurrences of $sf=7$ is below a value counted by the variable $NUM_{sf}(7)$. Then it passes to $sf=8$ and so on. The vector $NUM_{sf}$ counts the number of times the $sf$ is assigned in the system. Notice that the objective is to have a percentage of nodes at the different SFs with the final distribution as in table \ref{tab:airtime}.
In the phase two, we derive all the consecutive $ED_{GW}$ vector elements that have different coverage GWs, in terms of number of covered gateways and gateways identifier. These couple of nodes will be set with the same SF. Also in case of concurrent transmission, both the packets are correctly received from different gateways, and they will be set with the same SF value. In the figure, $sf=7$ is assigned to the two consecutive nodes with the respective GW coverage identifier values of $2,3$ and $2,4,5$.
In the third phase, we set the SF values of the remaining nodes according to the probability distribution function for the EXPLoRA water filling, except for the EDs that are already allocated. The first operation of the phase 3 is to update the distribution function by considering the already assigned nodes to the different SFs. 
The presented algorithm differs from the real implementation only for a further check before the SF assignment. If the correct reception is not guaranteed, due to the low $RSSI$ value, the real implementation forces the SF assignment only for a subset of SFs for which the correct reception is guaranteed. 

\section{Performance Evaluation}
\label{sec:performance_evaluation}
We evaluated the performance of EXPLoRa-C by using the LoRaSim simulation framework, in both the single cell and multi-gateway scenarios. All the simulation parameters are reported in table \ref{tab:Simul}, and channel capture is enabled with a conservative target $SIR$ of $1 dB$.
\vspace{-2ex}
\subsection{Single cell scenario}
\begin{figure}[t!]
\begin{center}
	\subfigure[]{\includegraphics[width=0.38\columnwidth]{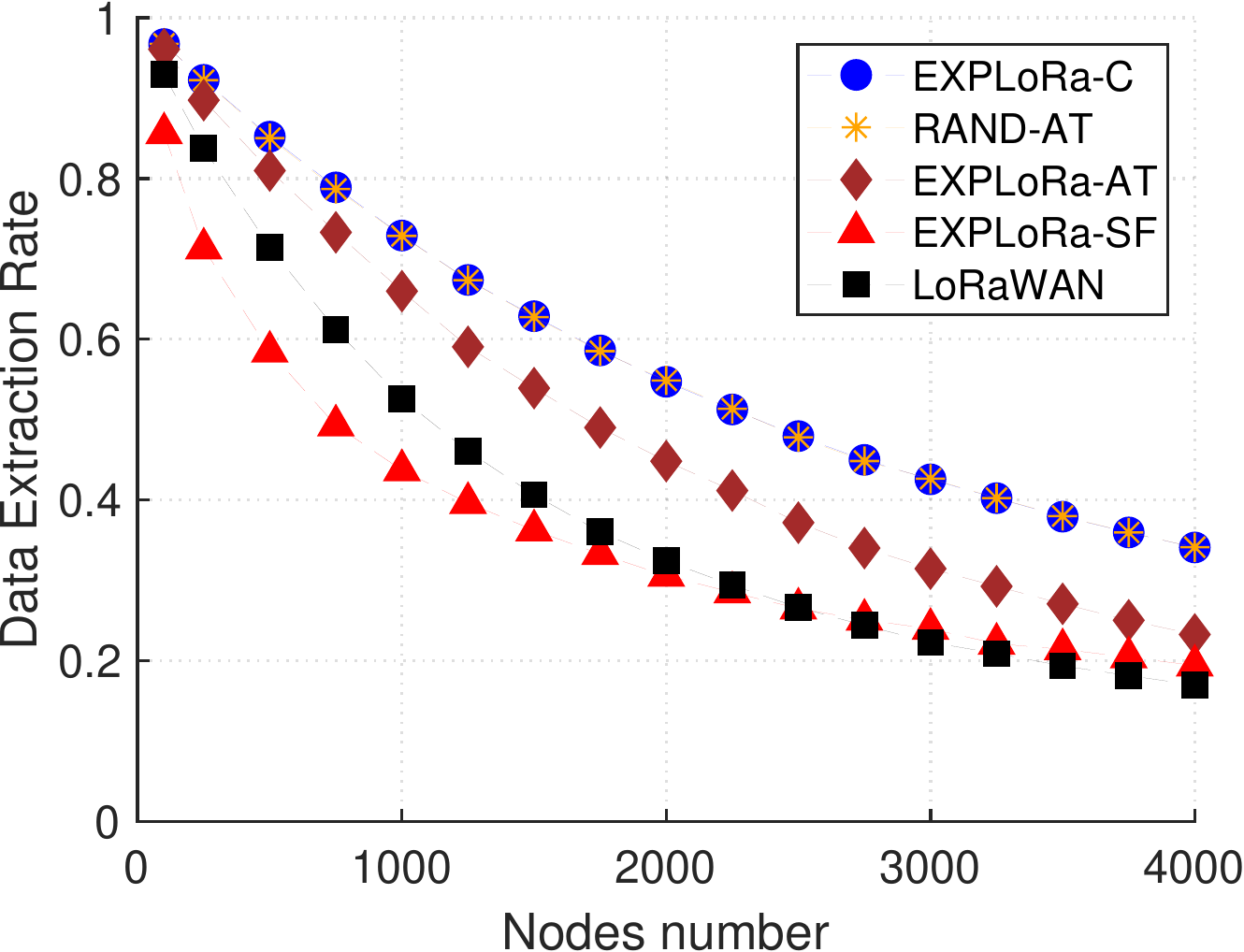}}
	\subfigure[]{\includegraphics[width=0.38\columnwidth]{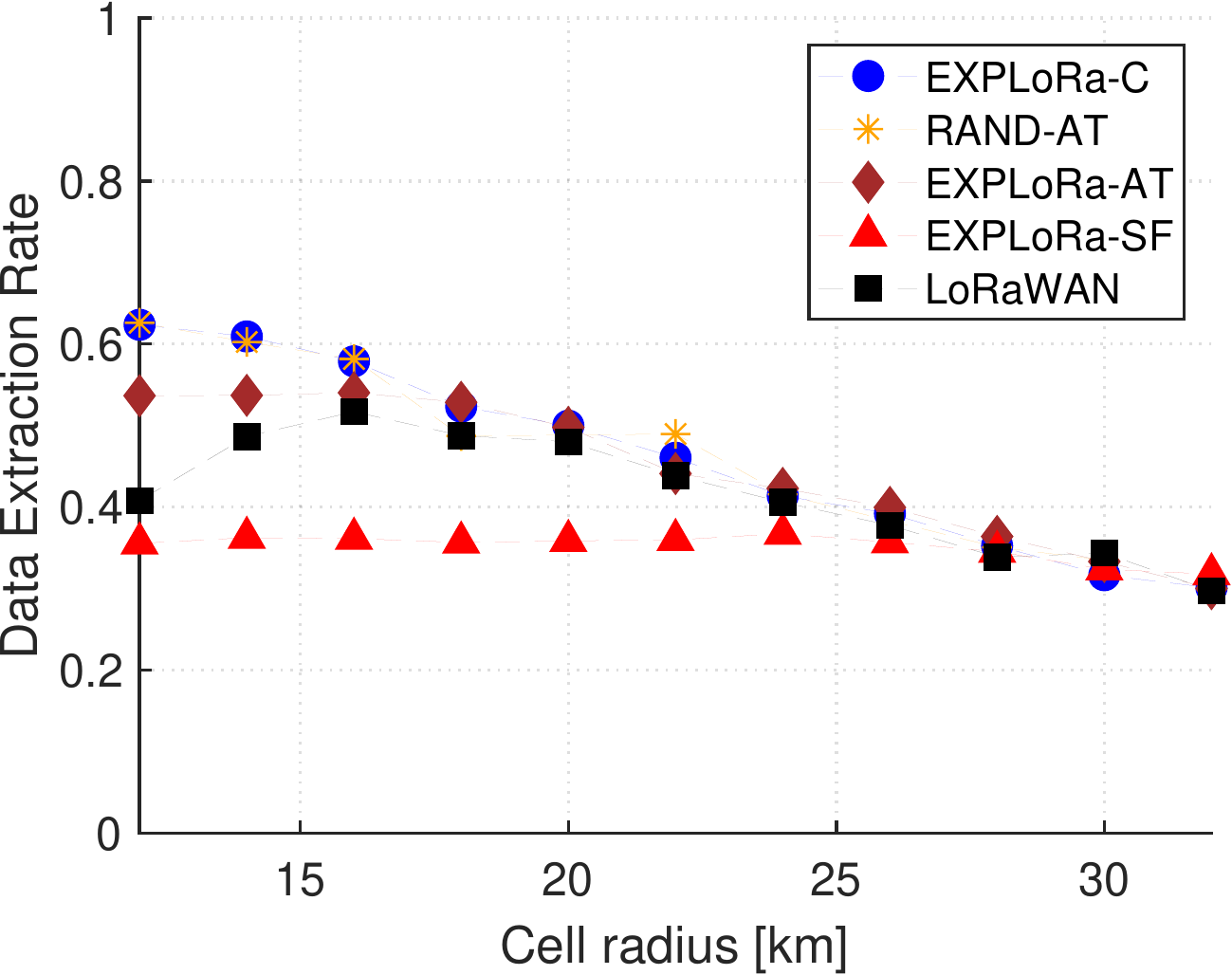}}
	\subfigure[]{\includegraphics[width=0.38\columnwidth]{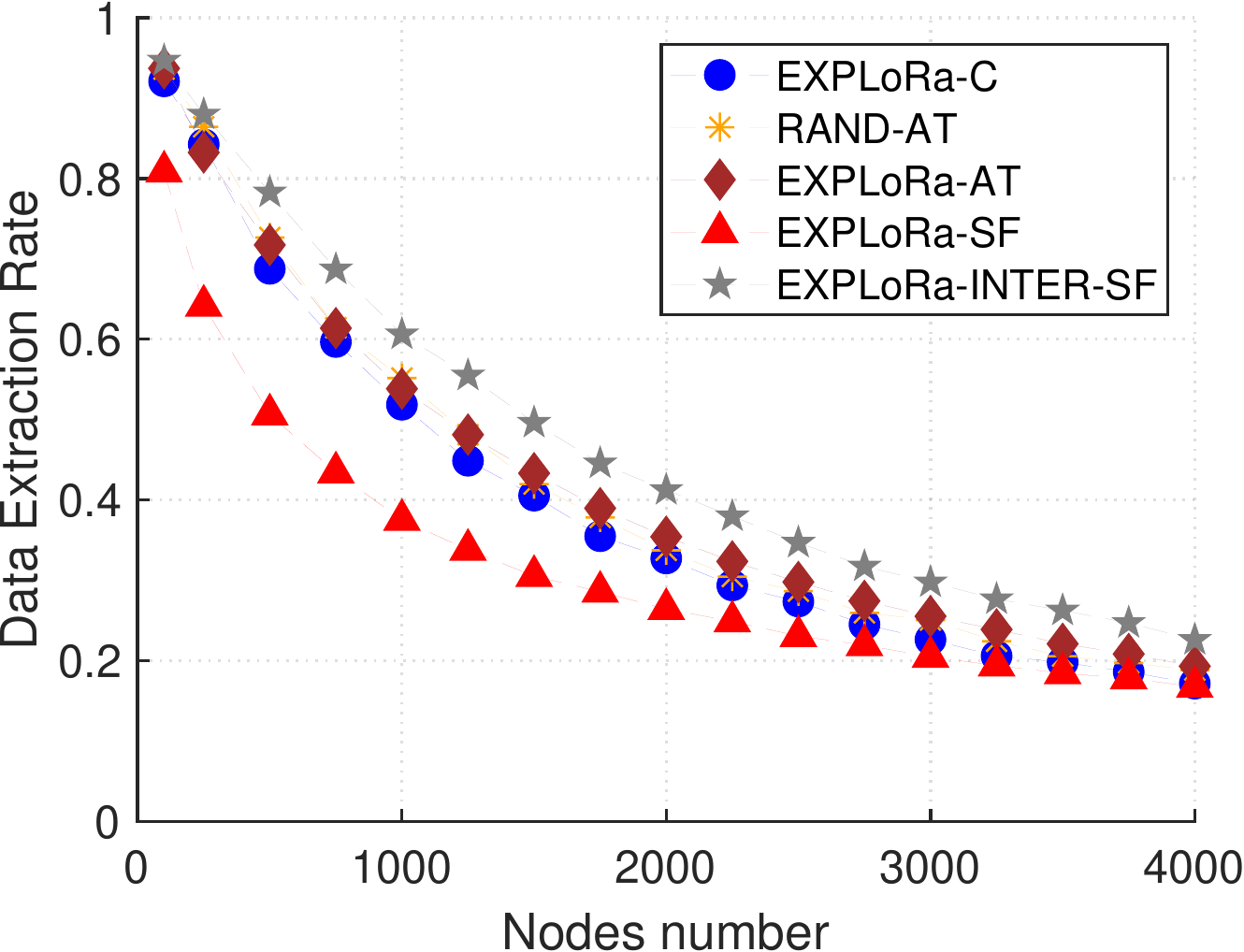}}
	\caption{$DER$ as a function of the EDs numbers in presence of channel capture, comparison among EXPLORA-C, RAND-AT, EXPLoRa-AT, EXPLoRa-SF and LoRaWAN, when cell radius is $12\;km$ (a). DER as a function of the distance in presence of channel capture when the maximum cell radius is $34\;Km$ (b). Simulation results in case of inter-SF interference when cell radius is $12\;km$ (c).}
	\label{fig:DER_1cell}
\end{center}
\vspace{-6ex}
\end{figure}
We considered two cell dimensions, namely a radius $R$ of $34\;km$ and $12\;km$ respectively. The former to represent scenarios in which the link budget is constrained as for SF allocations. The latter instead is an unconstrained deployment. Indeed, for the unconstrained deployment any node can use any SF, because with the considered propagation model and transmission power entails a $RSSI$ at the cell edge that is enough for using $sf=7$, i.e. the highest rate. We compare the results obtained by performing completely random allocations or by using different variants of EXPLoRa and the LoRaWAN legacy scheme.
Figure \ref{fig:DER_1cell}-a plots the average $DER$ achieved in the unconstrained deployment and a varying number of EDs (from 100 to 4000), which are uniformly placed within the cell area. We first assume that SFs are perfectly orthogonal and do not interfere each other. From the figure, we can observe that just equalling splitting the EDs across all the available SFs is not a good strategy: indeed, EXPLoRa-SF achieves performance which are worse or almost equivalent to the LoRaWAN legacy scheme. This is due to the load experienced on $sf=12$, which can reach unstable conditions even with a few hundreds of EDs. Specifically, when $s=1 pk/90\;sec$, the normalized load offered on $sf=12$ by each node is equal to 0.0132 ($ToA_{12}=1.19\;sec$) and therefore it is enough that $n_{12}$ is equal to 40 nodes (i.e. the total number of EDs $N$ is equal to  $6 \cdot n_{12} = 240$) to work in unstable conditions. In the case of LoRaWAN, since the link budget is not a constraint, all the EDs are always configured for working with $sf=7$, with a waste of cell capacity. Conversely, EXPLoRa-AT is able to optimize such a capacity, by equally sharing the normalized offered load on each available SF. However, by also optimizing the possibility to achieve channel captures in case of collisions, EXPLoRa-C can further improve the average $DER$, especially in high load conditions. Since in this scenario all the nodes can be served at the highest data rate, the performance achieved under a completely random SF assignment (named RAND-AT in the following) which obeys to the $P_{sf}(sf)$ proportions (i.e. randomly chooses $P_{sf}(sf) \cdot N$ nodes for using a given SF) are equal to the ones achieved with EXPLoRa-C.   
Indeed, when the number of nodes is high and nodes are placed in random way, the effect of phase 1 of the algorithm \ref{algo:explora-C} is limited: if the $RSSI$ values vary in a interval of $87 dB$ (from -$50 dBm$ to $-137 dBm$), only 87 nodes can be selected with a distance of at least $1 dB$ from the previous allocation. Therefore, most of the allocations are decided by the third phase of the EXPLoRa-C algorithm, which performs random choices according to the $P_{sf}$ proportions. 
In order to also take into account the effects of the link budget constraints, figure \ref{fig:DER_1cell}-b plots the $DER$ results achieved for a cell with $N=1500$, as a function of the cell radius (up to $34km$). Obviously, as the cell radius increases, the number of EDs which can make a choice on the adopted data rate gets smaller and therefore all the schemes tend to provide the same performance (i.e. the only possible choice is selecting the highest possible data rate compatible with the available link budget). The benefits of EXPLoRa are maximized for unconstrained deployments. Another factor to be considered is the interference between different SFs. Figure \ref{fig:DER_1cell}-c shows the $DER$ performance for a cell of $12km$ of radius, in presence of inter-SF interference. Performance are degraded for all the schemes, because of the increased competing load offered by EDs configured on different SFs. However, using the new load balancing criterion provided in equation \ref{e:load2} allows to optimize the performance, as depicted in the EXPLoRa-INTER-SF curve.  
\begin{figure}[t!]
\begin{center}
	\subfigure[]{\includegraphics[width=0.38\columnwidth]{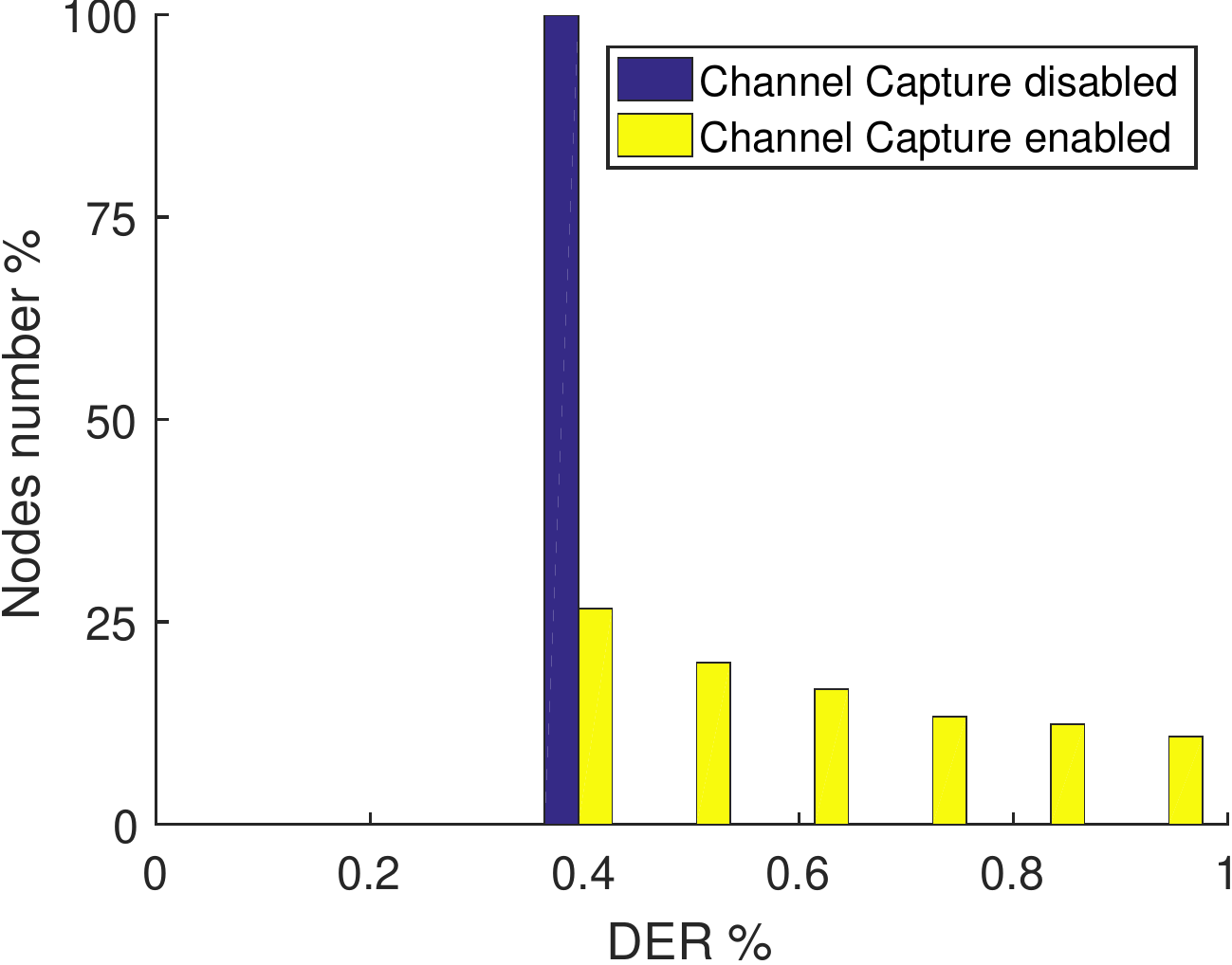}}
    \subfigure[]{\includegraphics[width=0.38\columnwidth]{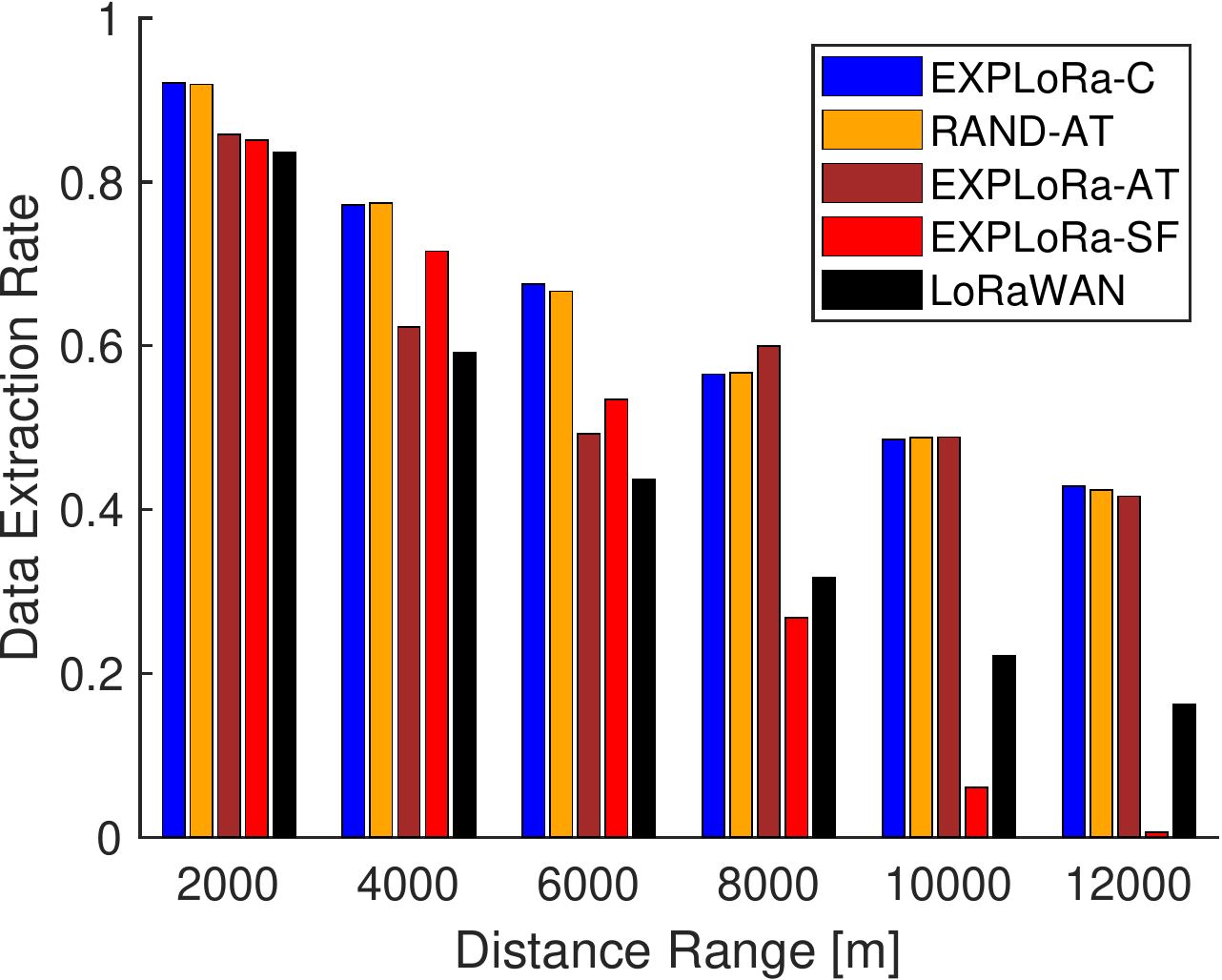}}
	\subfigure[]{\includegraphics[width=0.38\columnwidth]{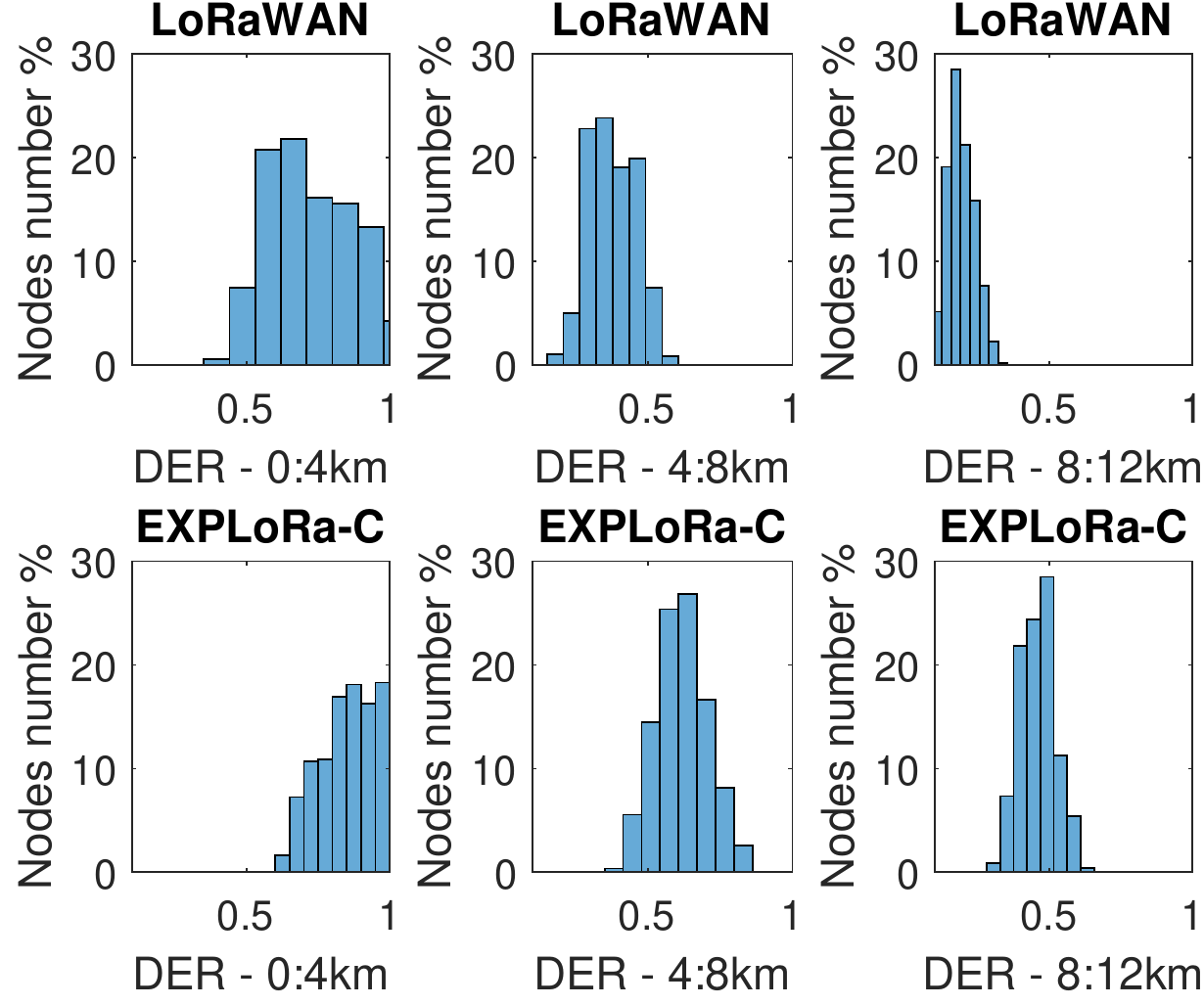}}	
	\caption{
	Histogram of DER for 750 at $sf=7$ (a). DER as a function of distance, relative to 6 circula rings, comparison among EXPLORA-C, RAND-AT, EXPLoRa-AT, EXPLoRa-SF and LoRaWAN (b). Histogram of the $DER$ achieved by the nodes for two schemes in 3 circular rings (c).}
    \label{fig:sf_lorasim_explora_comparison_maximum_cell_radius}
\end{center}
\vspace{-6ex}
\end{figure}
We could argue that boosting the capture effects can result in unfair performance between EDs. However, we should consider that the capture effect improves the $DER$ experienced by some nodes without degrading the performance of the other ones involved into the collisions. Figure \ref{fig:sf_lorasim_explora_comparison_maximum_cell_radius}-a shows the histogram of the $DER$ achieved by 750 nodes configured at $sf=7$, when  channel capture  is disabled (blue bar) or enabled (yellow bar). Without capture, all collisions result in multiple packet losses and all the nodes experience the same performance; in presence of capture, nodes closer to the gateway can take advantage of their physical position and have a successful transmission for some collisions with nodes placed at longer distances. 
To better present this result, figure \ref{fig:sf_lorasim_explora_comparison_maximum_cell_radius}-b plots the average $DER$ obtained under different allocation schemes for 6 not overlapped circular rings. Each bars group presents the mean $DER$ of all the nodes placed in the specific circular ring for all the schemes (e.g. the group bar labeled with $12 km$ is relative to the circular ring where $12 km$ and $10 km$ are the outer circle and the inner circle respectively). From figure \ref{fig:sf_lorasim_explora_comparison_maximum_cell_radius}-b we can observe that, with respect to the legacy LoRaWAN scheme, EXPLoRa schemes improve the performance in all the circular rings, without generating unfairness. For 3 circular rings, we also present, in figure \ref{fig:sf_lorasim_explora_comparison_maximum_cell_radius}-c, the histogram of the $DER$ achieved by the nodes for the two schemes, LoRaWAN (subplots above) and EXPLoRa-C (subplots below). From figure we notice that the EXPLoRa-C, respect to LoRaWAN, improves the number of nodes that get a DER great than 0.5, in all the circular rings.
\subsection{Multi-gateway scenario}
\label{sec:multi-gateway-scenario}

\begin{figure}[t!]
\begin{center}
	\subfigure[]{\includegraphics[width=0.38\columnwidth]{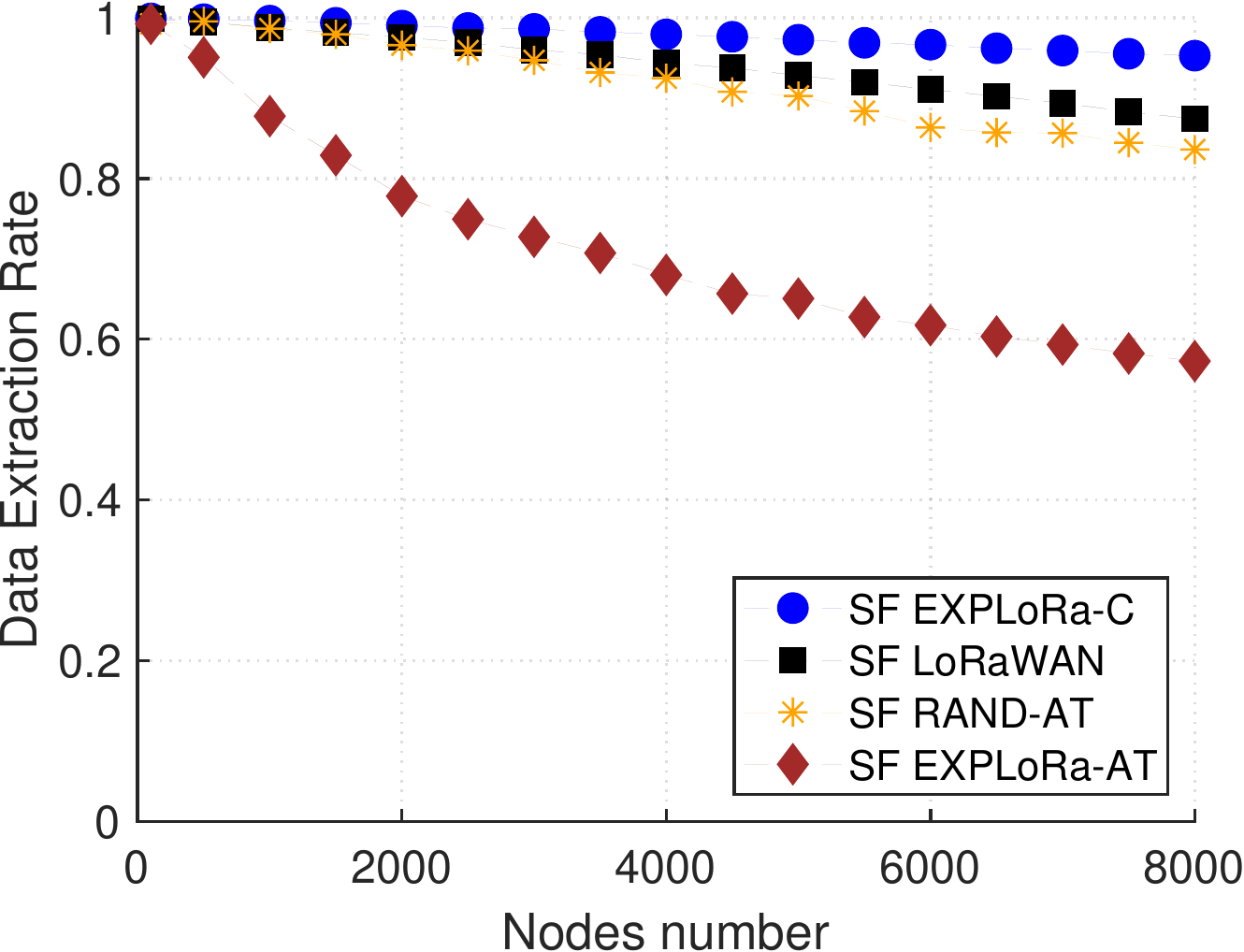}}
	\subfigure[]{\includegraphics[width=0.38\columnwidth]{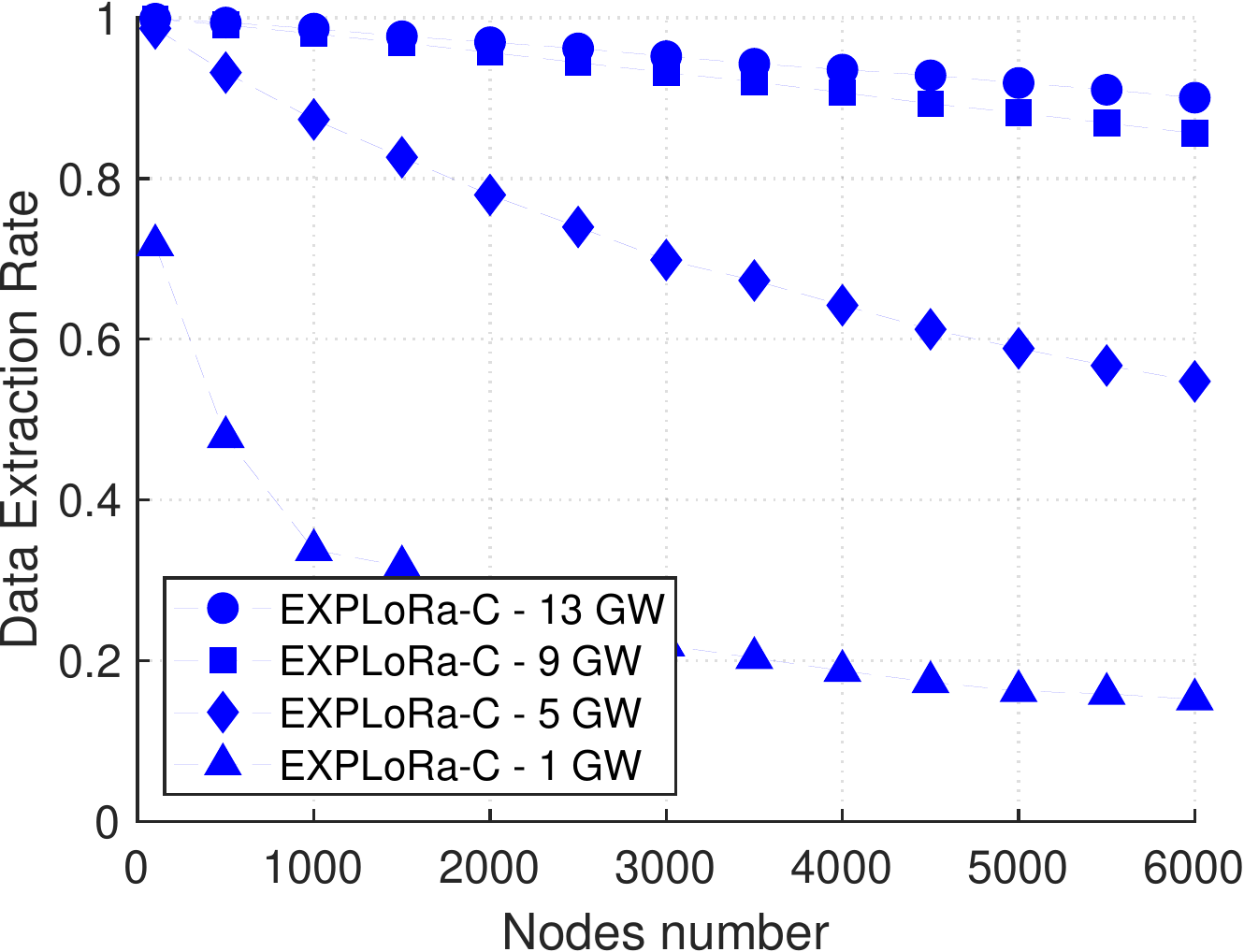}}	
	\subfigure[]{\includegraphics[width=0.38\columnwidth]{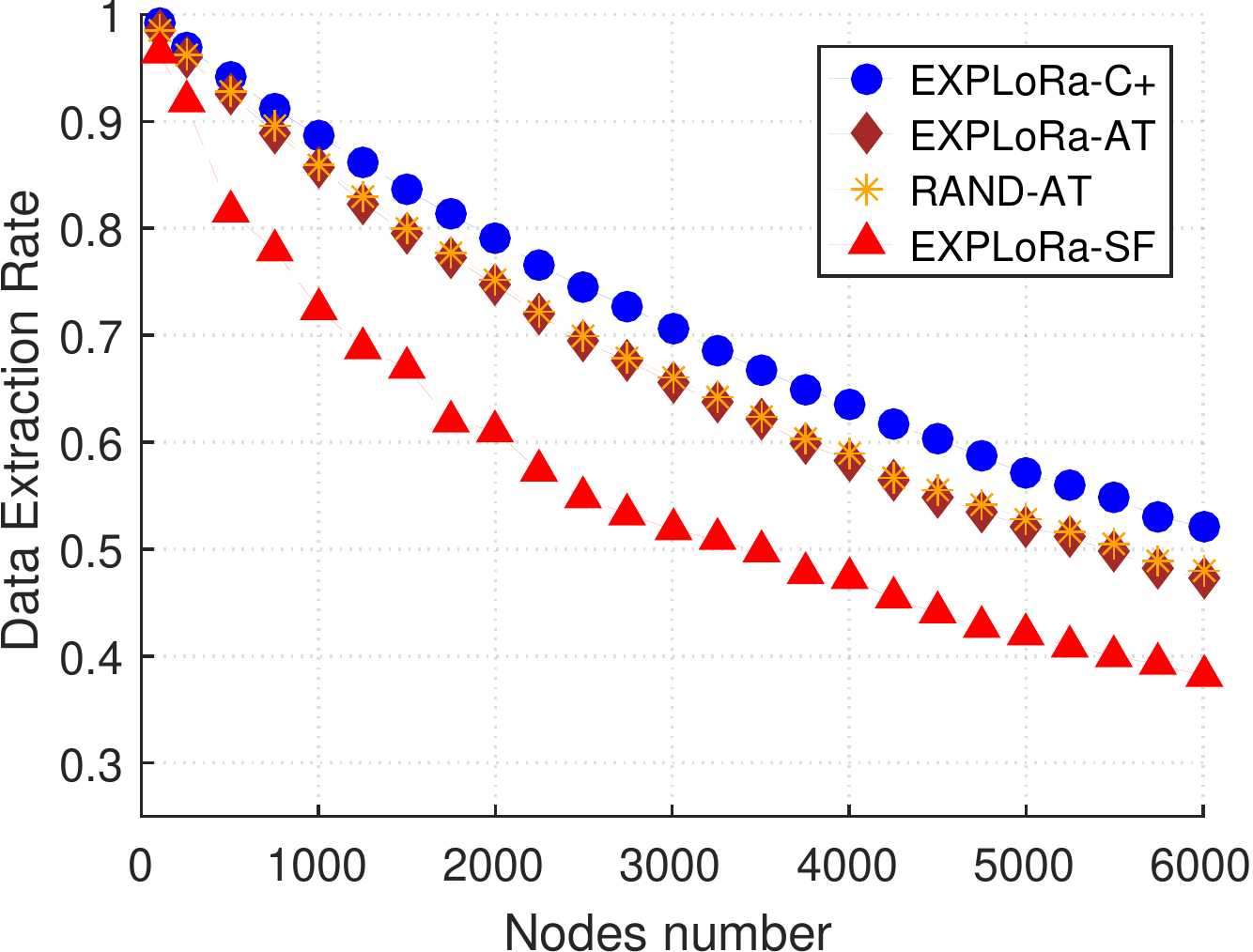}}
    \caption{Simulation results in multi-gateway scenario (a). EXPLoRa-C performance in multi-gateway scenario with different gateway number (b). EXPLoRa-C+ performance in multi-gateway scenario with coexisting operators (c).}
	\label{fig:multi}
\end{center}
\vspace{-6ex}
\end{figure}

For assessing EXPLoRa-C performance in a multi-gateway scenario, we considered two different network topologies: i) a regular grid, in which gateways are placed at regular distances and border effects are neglected; ii) a partitioned network with three gateways and border effects. 


The scenario corresponding to the regular grid is given by a topology with 25 gateways, regularly placed with a distance of $12\;km$ from the neighbors. 
The average $DER$ is presented in figure \ref{fig:multi}-a as a function of the number of EDs placed in the whole network (up to 8000 nodes). In this case, EXPLoRa-C provides not only the optimal results, but also a gain on the random allocations implemented by RAND-AT. The gain in comparison to EXPLoRa-AT is about $38\%$, and $8\%$ in comparison with legacy LoRaWAN ADR.

We also considered the capacity improvements that can be achieved by deploying an increasing number of gateways on a given area. In particular, for a circular area of $34km$ of radius, we performed other simulation runs with a number of gateways varying from 1 to 13. Figure \ref{fig:multi}-b shows the average $DER$ obtained when allocations are decided by EXPLoRa-C only. Increasing the number of gateways increases the capture probability, thus significantly boosting the overall system capacity even when the number of EDs is extremely high. 
Finally, figure \ref{fig:multi}-c  show the $DER$ values in the scenario of three coexisting operators with partially overlapped cells. 
in this case, it may happen that the interference perceived at each gateway from the neighbor cells could be biased towards some specific SFs, and therefore the load balancing between SFs is not guaranteed anymore by simply applying the $P_{sf}(sf)$ percentage independently on the customers of each operator. However, operators can estimate the interference provided by the other ones on each SF and reduce accordingly the load (as specified in equation \ref{eq:multi_operator}. This version of the scheme, called EXPLoRa-C+, provides the optimal $DER$ as shown in figure \ref{fig:multi}-c. 

\vspace{-2ex}
\subsection{Application in real scenario} 
\label{sec:realSceanrio}
In order to test the EXPLoRa-C scheme in a real LoRaWAN deployment, we considered a sensor network present in a small city in the north of the Italy. The network provides the water metering consumption of a 268 real consumer. Each water counter is equipped with a LoRaWAN module that forwards the real time measurement 18 times for week. The network is been deployed by an Italian operator, that also has installed 2 GWs to cover the whole region where the water meters are present. The packets, sent from the water meters, are forwarded by GWs to the operator NS. 
%
\begin{figure}[t!]
\begin{center}
	\subfigure[]{\label{fig:rssiDistrTc}\includegraphics[width=0.38\columnwidth]{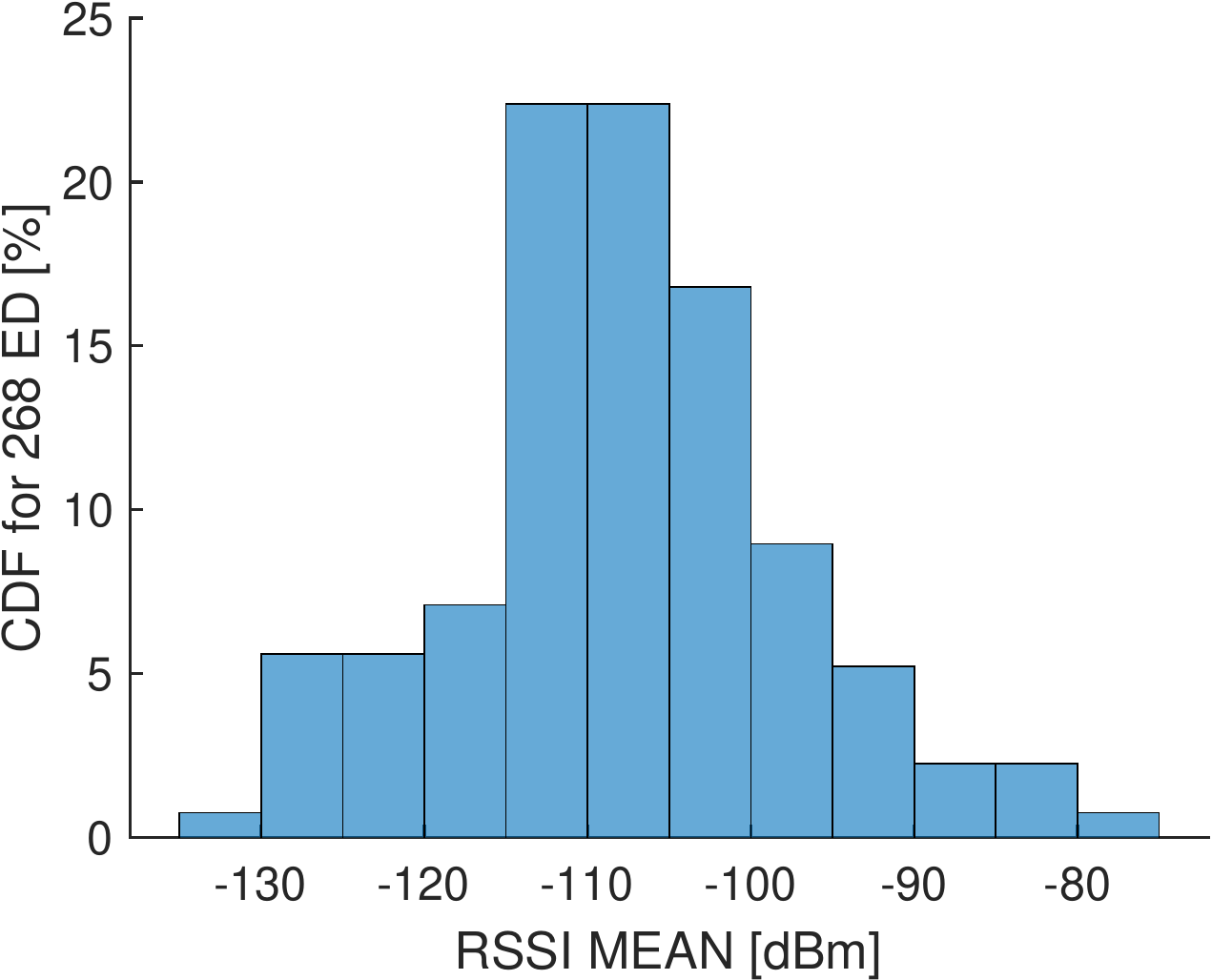}}
 	\subfigure[]{\label{fig:derTc}\includegraphics[width=0.38\columnwidth]{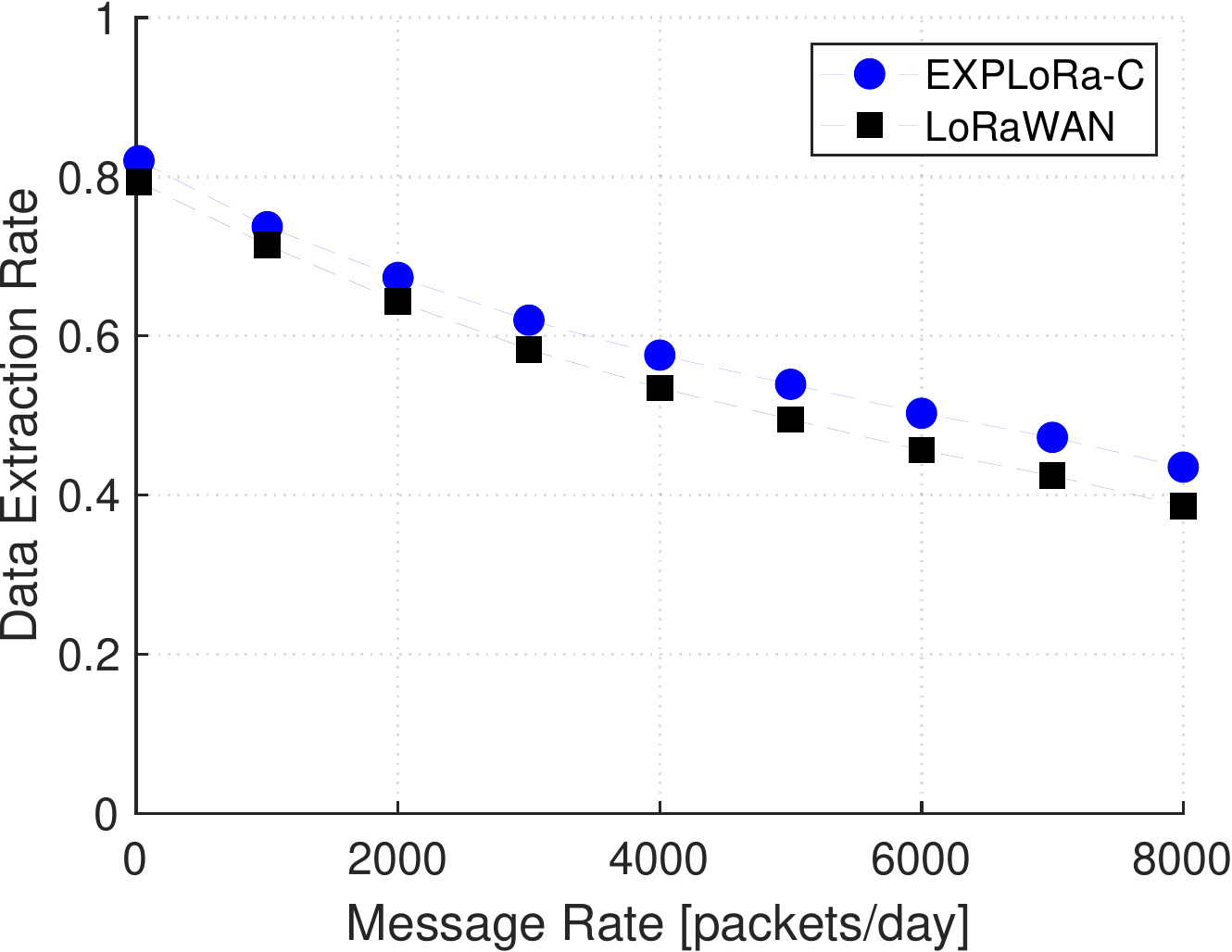}}	
	\caption{CDF of the $RSSI$ mean for the 268 devices present in the LoRaWAN network (a). Performance comparison in terms of DER for EXPLoRa-C and LoRaWAN schemes (b).}
	\label{fig:real-scenario-network-performance}
\end{center}
\vspace{-6ex}
\end{figure}
Each received packet is processed by the NS, the payload message is sent to the application service (provided by the public water manager), while the device radio link information are forwarded to the network controller. The network controller implements the legacy LoRaWAN ADR. All the received packets are also stored in a database, to allow future processing or anomaly detection. 
We use the database to study the behaviour of the devices in a range period of 6 months, specifically, we extract the $RSSI$ mean value for each device and SF assigned by the LoRaWAN ADR. Figure \ref{fig:real-scenario-network-performance}-a shows Cumulative Distribution Function (CDF) of the $RSSI$ for the 268 devices present in the network. We use the $RSSI$ mean and the SF values, assigned by the LoRaWAN ADR, to recreate the scenario by using the LoRaSim simulator. To reproduce the application scenario we also set a $\sigma^{2}$ variance of 6 to account the path loss shadowing.
Thus, we compare the performance of the EXPLoRA-C scheme versus the LoRaWAN ADR. Figure \ref{fig:real-scenario-network-performance}-b shows the DER for both EXPLoRa-C and LoRaWAN schemes when the message rate increase from 18 to 8000 packets per day. From the figure we can observe that EXPLoRa-C gets better performance than LoRaWAN, especially when the message rate increases. In case of a message rate of 8000 packets per day, EXPLoRa-C gets a DER of 5\% greater than LoRaWAN.

\section{Related Work}	
\label{sec:related}
Starting from early 2015, thanks to the LoRa Alliance \cite{Alliance} and the Semtech \cite{LoRa} products, the LoRa technology and the LoRaWAN network operators have gained momentum giving rise to 100 countries and more than 500 members involved in the LoRa exploitation. As a consequence, also the scientific literature has produced a number of papers related to the LoRaWAN systems, their performance and applications \cite{Mitigating16}\cite{georgiou2017}\cite{Mikha2016}\cite{reynders2017power}\cite{Onthecoverage15}.
The limits and the performance expectations for LoRaWAN have been studied in papers like \cite{limits, Performance17} and \cite{Mitigating16}.
The work in \cite{limits} provides an overview of the capabilities and limitations of LoRaWAN. The Authors discuss the limitations due to the imposed duty cycle that, for given packet $ToA$, force devices to be off for long periods. Apart a fine tuning of the SF allocation they also claim that the network deployment, e.g., the LoRaWAN gateways must be carefully dimensioned and planned to meet the requirements of each use case.
Voigt et al. in \cite{Mitigating16}, through simulations based on real experimental data, show the effects of the interference on performance of a LoRa network. 
Scalability issues in the LoRa system are analyzed in \cite{Bor2016}\cite{georgiou2017}\cite{Mikha2016}. Bor et al. in \cite{Bor2016} provide a LoRa link behaviour characterization by using practical experiments able to describe (i) communication range in dependence of communication settings of SF and Bandwidth (BW) and (ii) capture effect of LoRa transmissions depending on transmission timings and power. They also provided a LoRa simulator (LoRaSim) and evaluated the LoRa scalability limits in static settings comprising a single sink, and assessed how such limits can be overcome with multiple sinks and dynamic communication parameter settings. 
Scalability, has been also evaluated in case  of simultaneous transmissions using the same SF as well as different SFs in \cite{Mahmood18}. While, the co-SF interference is  natural  and  requires  $SIR$  protection  to  have  any  benefits from  capture  effect,  the  imperfect  orthogonality  among  SFs can also cause a significant impact in high-density scenarios.
However, in both cases a balancing of the load on the different SFs, as provided in this paper, can be of high benefit for the whole network.

	
	A measurement-based assessment of LoRa was carried out in \cite{Onthecoverage15}, which captures the $RSSI$ by different locations from the gateway and derives an heat map able to characterize performance as a function of the distance and of the environmental conditions (on water and on the ground). The paper also derives a channel attenuation model based on the presented measurements results. Empirical evaluations have been also provided in \cite{mikhaylov2017lorawan}.
	
	In \cite{georgiou2017} the effects of interference in a single gateway LoRa network have been investigated. Unlike other wireless networks, LoRa employs an adaptive chirp spread spectrum modulation scheme, thus extending the communication range in absence of any interference.
	Interference is however present when signals simultaneously collide in time, frequency, and SF. Leveraging
	tools from stochastic geometry, the authors of \cite{georgiou2017} have formulated and solved
	two link-outage conditions that can be used to evaluate the LoRA  behavior.
	
	The paper \cite{Mikha2016} analyzed the performance of the LoRa LPWAN technology by showing that following the current specification release, a single ED located close to the gateway can feature an uplink data transfer channel of only $2\:kbit/sec$ at best. In terms of scalability, they showed that a single LoRaWAN cell can potentially serve several millions of devices sending few bytes of data per day. Nonetheless, they showed that only a small portion of these devices can be located sufficiently far away from the gateway.
	Finally, \cite{magrinperformance} derives throughput behavior and capacity limits under some ideal conditions (perfect orthogonality of the SFs).


The recent literature also concentrated on the ADR mechanisms. These are designed to optimize the data rates, $ToA$ and energy consumption of the network devices \cite{Explora}\cite{Reynders17}\cite{Cionca18}. ADR should be enabled whenever an ED has sufficiently stable RF conditions and the idea is that, by setting the SF, a network controller can give higher data rates and radio visibility to specific EDs in the network.

In our preliminary work \cite{Explora} we provided a novel strategy, named EXPLORA-AT for implementing a suitable ADR in LoRaWAN systems. The key idea was to assign the SFs to the EDs in a way that assures an equal time on air occupation to all the available SFs
In this work, we consider the multi-gateway scenario and, we demonstrate that the EXPLoRa-AT performance can be further improved by take into consideration the channel capture. Differently from the heuristic provided in the multi-gateway case in \cite{ADMAIORA}, here we extend the the EXLORA-AT water filling scheme in a sequential way tailored for a multi-gateway scenario.

The work by Reynders et al. \cite{Reynders17} uses a genetic algorithm to accommodate in an optimal manner the 6 different SFs and 5 different power settings. In case of $N$ nodes the search space for optimal SF and power allocation per node is $N^{6\time5}$. The proposed scheme acts without considering the capture effect and by assuming orthogonal SFs,

Also the work in \cite{Cionca18} plays with the SF and the power allocation to achieve a fair adaptive data rate allocation, called FADR, tested in a single gateway scenario. They divide the area around the gateway in regions and assign, in accordance to the optimal distribution of \cite{Reynders17}, the SFs and the transmission power to control also the capture effect.

Finally, the channel  assignment  can be formulated also as a  many-to-one  matching game by treating LoRa users and channels as two sets of selfish players aiming to maximize their own utilities. This approach has been proposed in \cite{McCann} in the case of a single GW configuration. Tha approach works well when the Lora users are aware of the channels condition and occupation.


\section{Conclusions} 
\label{sec:conc}
In this paper, we have proposed and assessed an innovative ``sequential waterfilling'' strategy for assigning SF to EDs. Our driving intuition, duly backed up by theoretical results in section \ref{sec:capacity}, consists in equalizing the \emph{Time-on-Air} of the packets transmitted by the system's EDs in each spreading factor's group. This baseline idea has been embodied in a concrete algorithm, named EXPLoRa-C (where C stays for Capture), which is developed in a multi-gateway setting and takes into further consideration the channel capture effect, owing to its more significant role in LoRa with respect to other traditional wireless technologies.

Extensive simulation results for different loads and multi-gateway topologies show the improved effectiveness of EXPLoRa-C with respect to the currently employed ADR (Adaptive Data Rate) allocation algorithm, and show the robustness and adaptability of our approach to a wide range of scenarios.

We believe that a further asset of EXPLoRa-C consists in its implementation simplicity, which makes it appealing for integration in real world large-scale multi-gateway networks. As a matter of fact, our challenging ongoing work indeed consists in adapting EXPLoRa-C to operate in a real-world metropolitan-scale multi-gateway deployment in the Roma (Italy) area. 

	
\bibliographystyle{IEEEtran}
\bibliography{biblio}

\end{document}